\documentclass[final]{svjour3}
\usepackage{graphicx}
\usepackage{rotating}
\usepackage{amssymb}
\usepackage{mathptmx}
\usepackage[numbers]{natbib}
\usepackage{amsmath}
\usepackage{amsfonts}
\usepackage{placeins}
\usepackage{physics}

\newenvironment{psmallmatrix}
    {\left(\begin{smallmatrix}}
    {\end{smallmatrix}\right)}

\let\Oldsection\section
\renewcommand{\section}{\FloatBarrier\Oldsection}

\let\Oldsubsection\subsection
\renewcommand{\subsection}{\FloatBarrier\Oldsubsection}

\let\Oldsubsubsection\subsubsection
\renewcommand{\subsubsection}{\FloatBarrier\Oldsubsubsection}

\def\makeheadbox{\relax}

\bibpunct{}{}{,}{s}{}{,}

\begin{document}

\newcommand{\hdblarrow}{H\makebox[0.9ex][l]{$\downdownarrows$}-}
\title{Quantum noise  in carbon nanotubes as a probe of correlations in the Kondo regime}
\date{Published 2019}

\authorrunning{M. Ferrier et al.}
\titlerunning{Quantum noise  in carbon nanotubes}

\author{Meydi Ferrier$^1$, Rapha\"elle Delagrange$^1$, Julien Basset$^1$, H\'el\`ene Bouchiat$^1$, Tomonori Arakawa$^2$, Tokuro Hata$^2$, Ryo Fujiwara$^2$, Yoshimichi Teratani$^3$, Rui Sakano$^4$, Akira Oguri$^3$, Kensuke Kobayashi$^{2,5,6}$, Richard Deblock$^1$}

\institute{$^1$Laboratoire de Physique des Solides, CNRS, Univ. Paris-Sud, Universit\'e Paris Saclay, 91405 Orsay Cedex, France\\
$^2$Graduate School of Science, Osaka University, Toyonaka, Osaka 560-0043, Japan.\\
$^3$Department of Physics, Osaka City University, Sumiyoshi-ku, Osaka 558-8585, Japan\\
$^4$The Institute for Solid State Physics, The University of Tokyo, Kashiwa, Chiba 277-8581, Japan\\
$^5$Center for Spintronics Research Network (CSRN), Graduate School of Engineering Science, Osaka University, Machikaneyama 1-3, Toyonaka, Osaka 560-8531, Japan\\
$^6$Institute for Physics of Intelligence (IPI) and Department of Physics, Graduate School of Science, The University of Tokyo,Hongo 7-3-1, Bunkyo-ku, Tokyo 113-0033, Japan\\
\email{meydi.ferrier@u-psud.fr or richard.deblock@u-psud.fr}}

\maketitle

\begin{abstract}

Most of the time, electronic excitations in mesoscopic conductors are well described, around equilibrium, by non-interacting Landau quasi-particles. This allows a good understanding of the transport properties in the linear regime. However, the role of interaction in the non-equilibrium properties beyond this regime has still to be established. 
A paradigmatic example is the Kondo many body state, which can be realized in a carbon nanotube (CNT) quantum dot for temperatures below the Kondo temperature $T_K$. As CNT possess spin and orbital quantum numbers, it is possible to investigate the twofold degenerate SU(2) Kondo effect as well as the four fold degenerate SU(4) state by tuning the degeneracies and filling factor. 
Our article aims at providing a comprehensive review on our recent works on the Kondo correlations probed by quantum noise measurement both at low and high frequencies.
At low frequency, combining transport and current noise measurements in such a dot, we have identified the SU(2) and SU(4) Kondo states. Our experiment shows that a two-particle scattering process due to residual interaction emerges in the non-equilibrium regime. The effective charge $e^*$, which characterizes this peculiar scattering, is determined to be $e^*/e =1.7 \pm 0.1$ for SU(2) and $e^*/e =1.45 \pm 0.1$ for SU(4), in perfect agreement with theory. This result demonstrates that current noise can detect unambiguously the many-particle scattering induced by the residual interaction and the symmetry of the ground state.

In addition, we have measured the high frequency emission noise of a similar carbon nanotube QD in the Kondo regime, at frequencies of the order of $k_B T_K/h$. At the lowest measured frequencies the derivative of the noise exhibits an expected Kondo peak. However, this peak is strongly suppressed at higher frequency, pointing towards the existence of a high frequency cut-off of the electronic emission noise at a Kondo resonance. This leads us to postulate that a new timescale, related to the Kondo energy $k_B T_K$, emerges besides the timescale associated to the transport of electrons through the dot, given by the coupling to the reservoirs.

The goal of this overview article on our recent research is to demonstrate how current noise measurements yield new insight on interaction effects and dynamics of a Kondo correlated state. 


\end{abstract}

\section{Introduction}

The Kondo effect is a many body phenomenon arising, in condensed matter, when a localized quantum degree of freedom is coupled to a Fermi sea of delocalized electrons. It leads to the screening of this degree of freedom, manifesting as the formation of a resonance at the Fermi energy in the density of states for temperatures below the Kondo temperature $T_K$. The Kondo state was first observed in dilute alloys with magnetic impurities \cite{VandenBerg1962,Gruner1974}, but it appears as well in various systems, like heavy fermion compounds \cite{Hewson1993}, nanowires or 2DEG quantum dots (QD) \cite{Goldhaber98,Cronenwett98,Nigard00}. In this paper, we focus on the realization of the Kondo effect in a carbon nanotube (CNT) QD, which forms an artificial Kondo impurity when it is weakly coupled to source and drain electrodes \cite{Nigard00}. Low energy properties of the Kondo singlet are now well understood thanks to transport \cite{Laird2015} and shot noise \cite{Yamauchi2011,Delattre2009,Ferrier2016} measurements. But its out-of-equilibrium behaviour as well as its dynamics are theoretically still under investigation and experimentally almost unexplored. In this context we use current noise measurement to probe the out-of-equilibrium and dynamics of a carbon nanotube quantum dot in two complementary limits, at low and high frequency. At low frequency, noise measurements provide the effective charge of current carriers and their statistics. At high frequency, in the quantum regime ($h \nu \gg k_B T$), noise can be described in terms of exchange of photons of energy $h \nu$ between the quantum dot and
the noise detector. Depending on whether photons are emitted or absorbed by the source, one can distinguish emission or absorption noise and access the dynamics of Kondo screening at frequencies of the order of $k_B T_K/h$. 

In the low frequency regime, around equilibrium (voltage bias $V\approx 0$), electronic properties of such a dot in the Kondo regime are well described by non-interacting quasiparticles, in the spirit of Landau theory of Fermi liquid.  This allows  a good understanding of the transport properties in the linear regime using for example Landauer-Buttiker scattering theory \cite{Blanter2000} or Kubo formalism.
However, going further in the non-equilibrium regime by increasing the bias voltage, a residual interaction between quasi-particles \cite{MahanBook} gives non-linearities and scattering of multiple-particles which are beyond the previous theory and have still to be understood. In quantum dots, this phenomenon appears for example in the Kondo state or in the inelastic cotunneling regime of Coulomb blockade.
We will see that, although non-linear conductance is already a signature of interactions,  current noise, for its part, provides a powerful tool which can detect unambiguously the many-particle scattering induced by the residual interaction and provides a precise understanding of the link between non-linear noise, Kondo correlations and symmetry of the ground state. This was possible thanks to both the high accuracy of the noise measurement set-up in Osaka and the high quality of the CNT fabricated in Orsay, which reaches unitary limit of the Kondo effect (perfect transmission) in both symmetries $SU(2)$ and $SU(4)$. In addition,  the high Kondo temperature of our sample combined with the use of a dilution fridge of base temperature 20 mK allowed us to investigate the linear regime of the Kondo regime ($k_BT<eV \ll k_BT_K$) where the non-interacting Fermi liquid description holds.

The emission noise in the quantum regime ($h\nu\gg k_BT$) is related to the splitting of the Kondo resonance and decoherence mechanisms. Actually, when a bias voltage $V$, drives the Kondo state out-of-equilibrium, the peak in the DOS splits, leading to one replica at the Fermi energy of each reservoir \cite{Monreal2005, Vanroermund2010, Lebanon2001, meir1993}. Such a split Kondo resonance is predicted to give rise to a logarithmic increase of the noise at $eV=h\nu$ \cite{Moca2011,Muller2013}. Moreover the peaks in the DOS may be weakened by decoherence induced by inelastic scattering \cite{DeFranceschi2002,Leturcq2005}. However, this picture is valid only when the two electrodes are symmetrically coupled to the dot, participating equally to the formation of the Kondo singlet \cite{Pustilnik2004}. Otherwise the best coupled reservoir mainly participates in the Kondo state \cite{Kehrein2005}, such that the associated resonance in the DOS is more proeminent than the one pinned on the less coupled contact \cite{Lebanon2001}. In the very asymmetric case, the Kondo resonance associated with the best coupled contact stays very close to equilibrium such that there should be very weak decoherence induced by the bias voltage. The emission noise should then probe an equilibrium Kondo resonance, expected to be very different from the out-of-equilibrium one in the symmetric case.
Thanks to an on-chip detection, we measure the emission noise \cite{Basset2012b}, associated with the emission of photons at frequency $\nu$ during tunnelling through the Kondo impurity in both symmetries $SU(2)$ and $SU(4)$. For the latter symmetry we probe noise in a regime which is always with a frequency below $k_BT_K/h$ and cannot be reliably compared to theory. For the $SU(2)$ case we consider different contact asymmetries, going from a very asymmetric case to a nearly symmetric one. We interestingly find a frequency cut-off for emission noise for both the symmetric and asymmetric cases, which is not predicted by theory \cite{Moca2011,Muller2013,Rothstein2009,Hammer2011,Zamoum2016}.

This article constitutes a summation of our recent works \cite{Ferrier2016,FerrierPRL,Basset2012a,Delagrange2018,DelagrangePhD} on the current noise in the Kondo regime, which is organized as follows. Part 2 presents a brief theoretical overview on the Kondo effect in CNT. Part 3 describes our two different experimental setups to measure the noise at low frequency or high frequency. Part 4 presents the results for the shot noise in the low frequency regime. Finally, Part 5 is devoted to the high frequency regime.

\section{Kondo Effect in Carbon Nanotube Quantum dots}

\subsection{Kondo state around equilibrium}
The Kondo effect was discovered in metallic alloys containing magnetic impurities to explain the increase of resistance at low temperature and the quenching of magnetic susceptibility. Jun Kondo \cite{Kondo1964} explained it from the antiferromagnetic coupling between spin of impurity and the one of conducting electrons at the Fermi energy modelized by the Hamiltonian:
\begin{equation}
H_K=\sum\limits_{\vec{k},\sigma}\epsilon_{\vec{k},\sigma}n_{\vec{k},\sigma}-J_K\vec{S}.\vec{s(0)}
\label{Hkondo}
\end{equation}
with $\vec{S}$ the spin of the impurity and $\vec{s(0)}$ the spin of the conduction electrons on the impurity site.
 It was then fully confirmed by the renormalization group analysis of K. Wilson \cite{Wilson1975} which could cure non physical divergences at low temperature of the Kondo initial perturbative approach. It was demonstrated that a resonance in the density of states (DOS) appears below the characteristic temperature $T_K$ due to the formation of a many-body singlet state which enhances the diffusion amplitude and screens the spin of the impurity  (and eventually creates divergences in the perturbative approach).
 
 This characteristic energy is given by:
 \begin{equation}
 k_BT_K\propto e^{-\frac{1}{N_0\abs{J_K}}}
 \label{Tk_wilson}
 \end{equation}
where $N_0$ is the density of states. It can be interpreted as the binding energy of the singlet ground state. Since conduction electrons are freely moving, it is hard to capture them yielding this exponentially small energy, whereas for localized spin the singlet/triplet gap is given by the exchange energy $J$ \cite{Pustilnik2004,Wilson1975,Anderson1970}. 

By using quantum dots as controllable impurities, it is now possible to investigate locally such a many-body state which is formed between the localized electron on the dot and the conducting electrons in the leads\cite{Goldhaber-Gordon1998}. This controllability has allowed the discovery of peculiar Kondo states when additional electronic degrees of freedom are involved in the screening mechanism. It can even give rise to non Fermi liquids as recently observed \cite{Iftikhar2015,Keller2015}. However in this article, we will concentrate on one class of Kondo states called $SU(N)$\footnote{$SU(N)$ refers to the symmetry group of the Hamiltonian describing the system at low energy (after renormalization)} where electrons on the dot and in the leads share the same "hyperspin" with $N$ degenerate states. This happens in CNT quantum dots, since electrons carry a fourfold degenerate spin made from the composition of spin and valley (or orbital) quantum numbers. Hence,  $SU(2)$ and $SU(4)$ Kondo states can be realized\cite{Jarillo-herrero2005}. The initial Kondo problem corresponds to the symmetry group $SU(2)$ where conduction electrons screen the spin 1/2 of the impurity. The $SU(N)$ Kondo model for a single-electron impurity with a magnetic moment degenerate $N$ times  is known as the Coqblin-Schrieffer model \cite{Coqblin1969}.

\subsection{Kondo state in quantum dots}
A quantum dot is an island on which the number of electrons is fixed and experimentally controllable. The conduction electrons are provided by the nearby leads connected to the dots. Depending on the dot/lead coupling, Kondo state emerges. Such a system can be described by the Anderson impurity model.

\begin{equation*}
H=H_{dot}+H_{lead}+H_{coupling}
\end{equation*}
\begin{itemize}
\item$H_{lead}=\sum\limits_{\vec{k},\sigma}\epsilon_kc^{\dagger}_{\vec{k},\sigma}c_{\vec{k},\sigma}$
\item$H_{dot}=\epsilon_d\sum\limits_{\sigma}d^{\dagger}_{\sigma}d_{\sigma}+U\sum\limits_{\sigma<\sigma'}n_{d\sigma}n_{d\sigma'}$
\item$H_{coupling}=\sum\limits_{\vec{k},\sigma}V_{\vec{k}d}a_{\vec{k},\sigma}d^{\dagger}_{\sigma}+hc$
\end{itemize}
Here $\sigma$ is a $N$-times degenerate hyper-spin. The dot-lead coupling\footnote{With this convention, $\Gamma$ corresponds to the width of the resonance i.e $\Gamma_L+\Gamma_R$ for a quantum connected to two leads.} is given by $\Gamma=2\pi N_0|V_{kd}|^2$.

If $\sigma$ represents a doubly degenerate spin $1/2$, we obtain the $SU(2)$ Kondo effect.
In the limit of a singly occupied dot ($\epsilon_d+U\gg\epsilon_F$) with weak charge fluctuations $(\Gamma\ll U)$ this Hamiltonian is equivalent \cite{Schrieffer1966} to the Kondo Hamiltonian in Eq. \ref{Hkondo} where $J_K$ is expressed as:
\begin{equation*}
J_KN_0=\frac{\Gamma U}{\pi\epsilon_d(\epsilon_d+U)}
\end{equation*}
which yields using Eq. \ref{Tk_wilson}:
\begin{equation}
k_BT_K=\frac{\sqrt{U\Gamma}}{2}e^{\frac{\pi\epsilon_d(\epsilon_d+U)}{\Gamma U}}
\label{Tk}
\end{equation} 
In this transformation, since the charge fluctuations are weak, the dot with a fixed number of electrons is replaced by a pure spin impurity $\vec{S}$ in the Kondo Hamiltonian. The tunneling events of electrons between dot and leads have been incorporated in an effective exchange coupling.
The ground state of this system is a many body singlet formed between electron on the dot (or the spin impurity) and conducting electrons in the leads (Fig. \ref{SU4GS} left). Its thermodynamics and transport properties can be calculated from the low energy excitations which can be described by the Landau theory of Fermi liquid as explained in paragraph 2.4. The main point is to calculate the phase shift experienced by an incoming electron, which is scattered by this many body state.

\subsection{CNT and SU(4) symmetry}
Due to their specific bandstructure, electrons in CNT present two degrees of freedom \cite{Laird2015}: the Zeeman spin $\sigma=\uparrow\ \downarrow$ and the valley isospin $K$ and $K'$ (or orbital momentum) which corresponds roughly to clockwise and anti-clockwise wavefunctions. Consequently, single particle states in CNT are fourfold degenerate and can be labelled as $\ket{K\uparrow}$, $\ket{K\downarrow}$,$\ket{K'\uparrow}$ and $\ket{K'\downarrow}$. This degeneracy is lifted by spin-orbit interaction due to the curvature or valley-mixing which comes from finite length or disorder in the CNT \cite{Marganska2015}.

\begin{figure}[h] 
\centering 
\includegraphics[width=0.28\columnwidth]{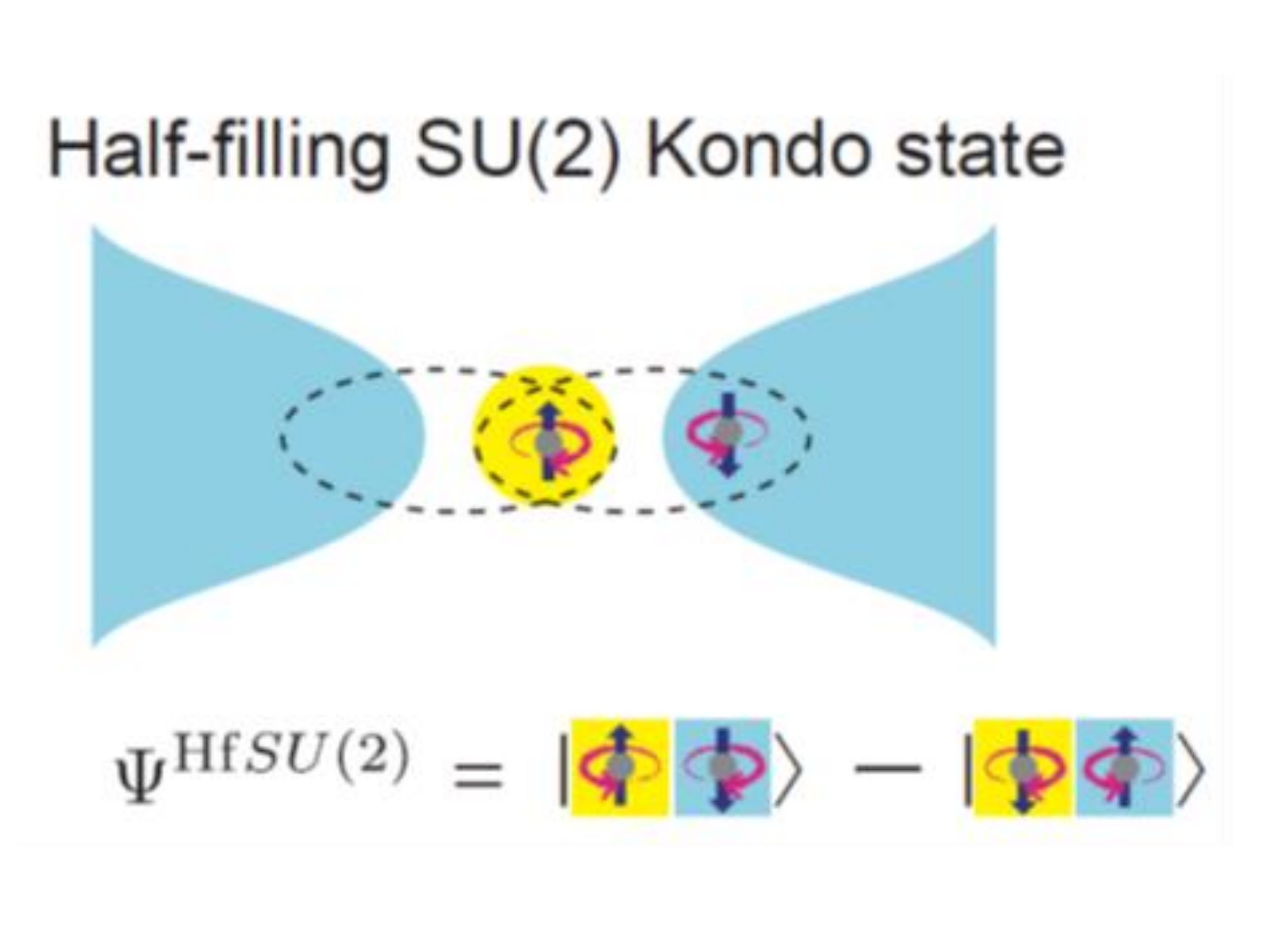}
\includegraphics[width=0.35\columnwidth]{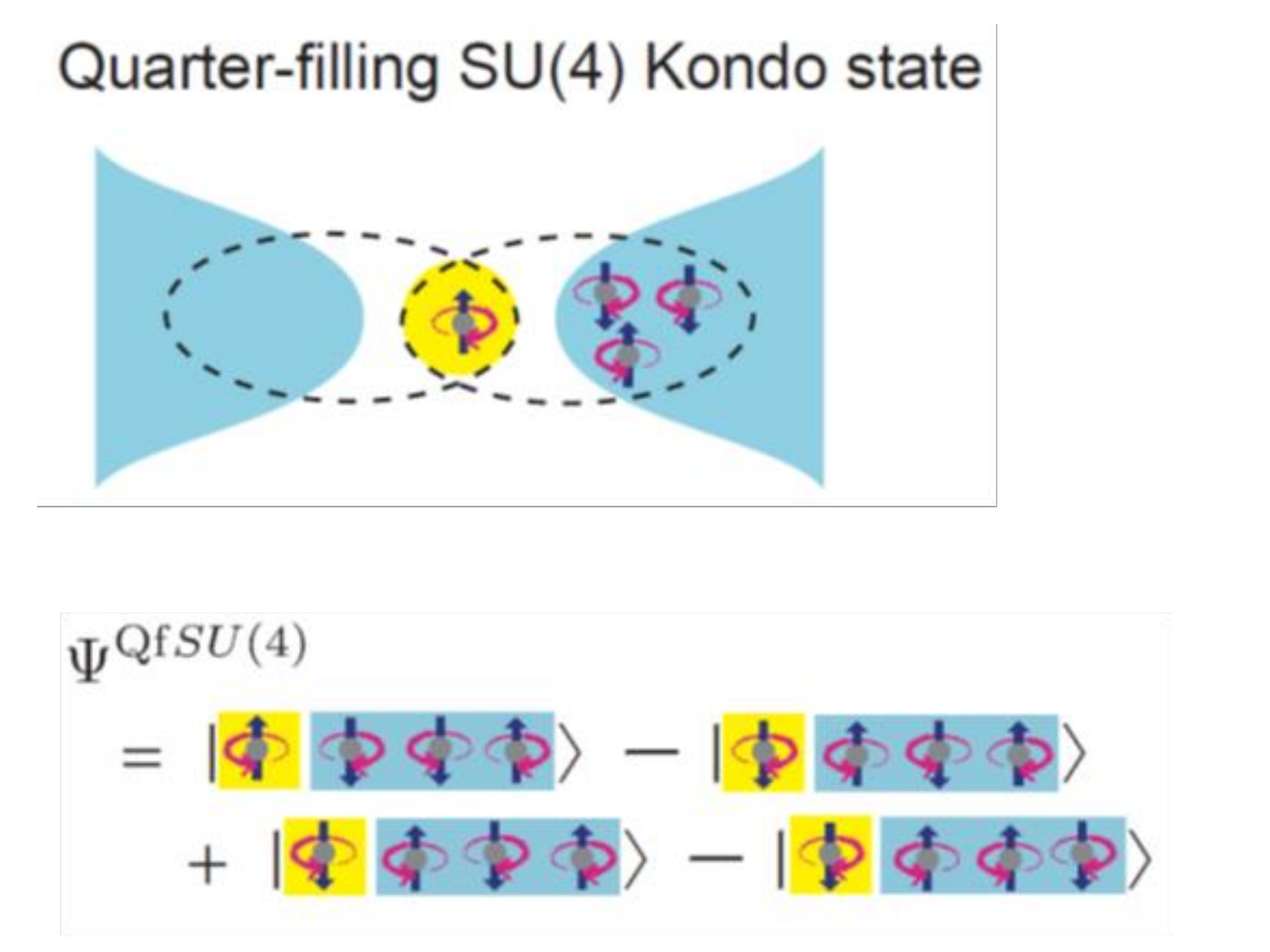}
\includegraphics[width=0.35\columnwidth]{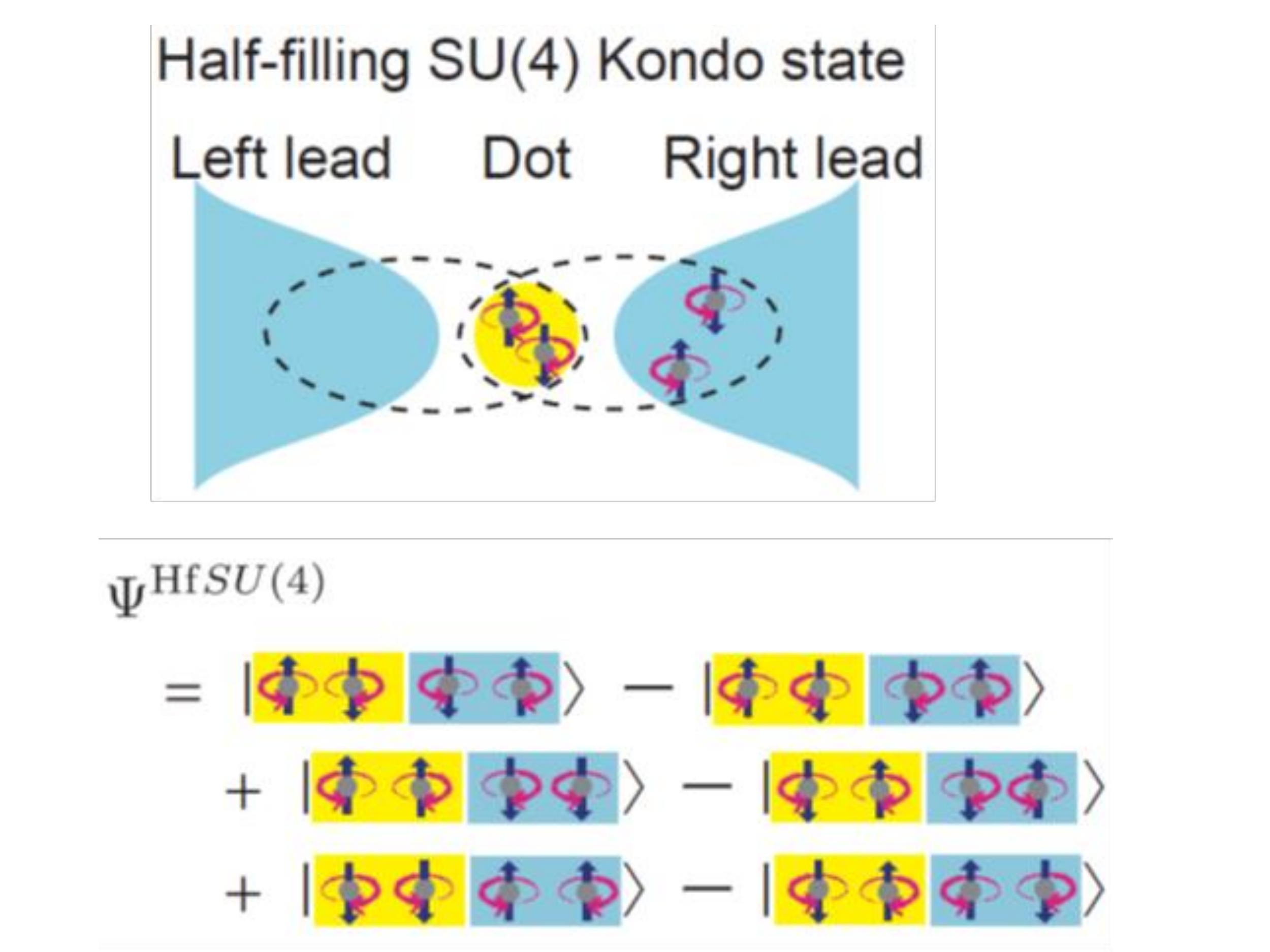}
\caption{Ground state of the Kondo effect in the different symmetries. Upper part: Scheme of the screening cloud between dot and leads for SU(2) symmetry at half-filling, SU(4) symmetry at quarter filling and SU(4) symmetry at half-filling ($N=2$). Lower Part: Ground state wave function in each case. Circular arrows represents the valley degree of freedom ($K$ and $K'$), vertical arrows represent the physical spin. Yellow represents electron on the dot, blue is for conduction electrons in the contact. Note that orbital degree of freedom is conserved in the contact.}
\label{SU4GS} 
\end{figure}

That is why two symmetries of the Kondo state are observed in CNT quantum dot \cite{Jarillo-herrero2005,Makarovski2007}. If valley degeneracy is lifted with a  gap larger than the characteristic scale $k_BT_K$, only  $\ket{K\uparrow}$ and $\ket{K\downarrow}$ states participate to the Kondo resonance and the usual $SU(2)$ symmetry is observed. On the other hand, in clean CNT, the four states are degenerate and participate to the Kondo resonance yielding the $SU(4)$ symmetry. In this case, Kondo effect occurs for odd numbers of electrons in the dot (m=1 or m=3) but also at half filling($m=2$). Indeed, at half-filling, $6$
 two-particle states are degenerate and can form a Kondo singlet ground state (see Fig.\ref{SU4GS}).

\subsection{Fermi liquid description of the Kondo state}
\label{Fermiliquid}
To go further and investigate thermodynamics and transport properties in the strong coupling regime ($T\ll T_K$), it was shown by Nozieres \cite{Nozieres1974} in the SU(2) symmetry and then generalized for SU(N) \cite{Mora2009b} that Kondo state can be described as a local Fermi liquid. 

In the spirit of Landau theory, Nozieres has explained how response functions (susceptibility, resistivity) evolve around equilibrium in the Kondo regime. In the limit of $T=0$, the renormalization group tells us that the ground state is the one corresponding to $J\rightarrow\infty$: a non-magnetic singlet state (see Fig. \ref{SU4GS} left: one conduction electron screen the impurity spin)  surrounded by a Fermi sea of free electrons. The singlet is frozen since $J=\infty$. In a tight binding picture, one site of the conducting electrons is removed to form the singlet state and decoupled from other sites.
To understand the transport properties, we have to think how incoming electrons from the leads are scattered by this singlet state. At equilibrium, considering s-waves for conduction electrons,  it can be reduced to a $1d$ problem where the scattering is encoded in the phase shift of the incoming wave. This phase shift is given by the Friedel sum-rule which yields for this Anderson's impurity model \cite{Langreth1966}:
\begin{equation*}
  \sum_{i=1}^{N/2}\delta_{0,i}=\frac{\pi}{2}m
\end{equation*}  
where $m$ is the number of  electrons displaced by the impurity and $i$ the number of conduction  channels\footnote{The elctron on the dot itself is not counted but only the electron of the conduction band which is needed to form the singlet state}. 
\subsubsection{SU(2) phase shift}
If electrons possess a two-fold degenerate degree of freedom (like usual spin), transport occurs through a single channel. The dot is singly occupied and one electron of the lead is displaced to form the singlet.
Hence we obtain $\delta_0=\frac{\pi}{2}$ for $SU(2)$.
In this case the problem can be seen as a 1d chain. The electron on the last site in contact with the dot is used to form the singlet, shifting the boundary condition of electronic waves by one site i.e a phase shift of $\pi/2$.
  
\subsubsection{SU(4) phase shifts}
  
In CNT, the SU(4) Kondo effect appears with the ground states depicted on Fig.\ref{SU4GS}. Transport occurs here through two conduction channels. At quarter filling ($m=1$ or $3$ electrons in the dot), an average number of one electron of the lead is needed to  form the singlet. For $m=2$, $2$ electrons participate to the ground state. Hence, the Friedel sum rule yields:
\begin{align*}
  \text{m=1} & & \delta_1=\delta_2=\frac{\pi}{4}
\end{align*}
\begin{align*}
  \text{m=2} & & \delta_1=\delta_2=\frac{\pi}{2}
\end{align*}
  
  \subsubsection{Effective interaction mediated by singlet-triplet fluctuations: phase shift near equilibrium}
  \label{STeffective_interaction}
  
  \begin{figure}[h] 
  \centering 
  \includegraphics[width=0.8\columnwidth]{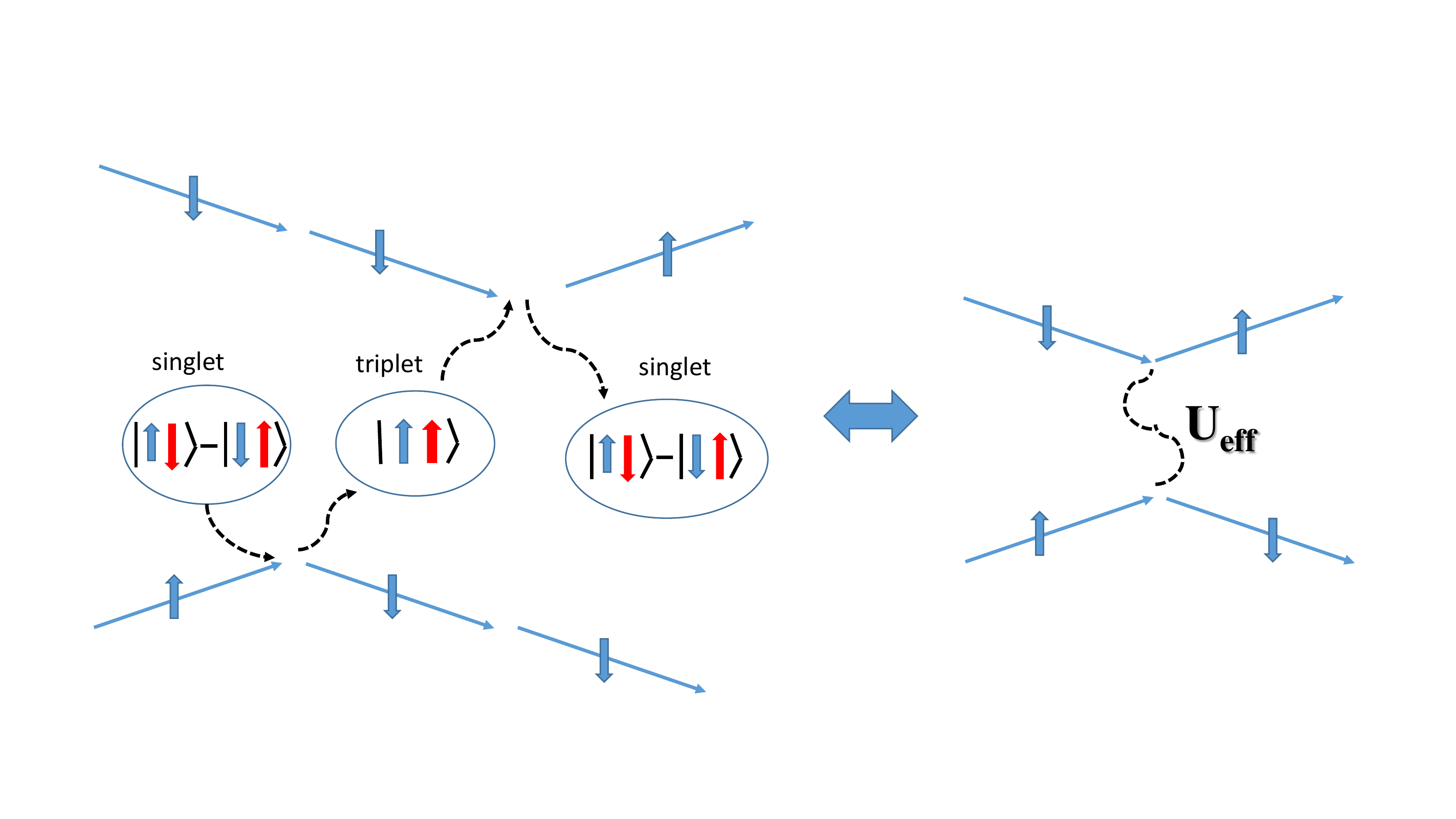}   
  \caption{Rough schematic view of the effective interaction mediated by the Singlet-Triplet transition of the Kondo Ground state. Red spin represent impurity spin. Blue spin is for conduction electrons. Two conduction electron of opposite spin can effectively interact through the Kondo impurity and exchange their spin.}
  \label{effective interaction} 
  \end{figure}
  
  To compute electronic properties beyond equilibrium ($T,V\geq 0$), we have to calculate the phase shift when J decreases from $\infty$. When $J$ decreases, virtual excitations of the singlet to the triplet state become possible which creates an effective interaction between two conducting electrons of opposite spin (like the exchange of a virtual phonon also creates an effective electron/electron interaction). This is schematically explained in Fig. \ref{effective interaction}. By looking at the scattering phase shift induced by  this effective interaction using Landau's prescription, Nozieres \cite{Nozieres1974} could derive the different susceptibilities and in particular conductivity or current in the linear response regime. As expected for a Fermi liquid, the conductivity is thus quadratic with $T$: $\sigma\propto\left(\frac{T}{T_K}\right)^2$, where $T_K$ is the only energy scale in the problem.
  
This approach shows that at low temperature, the low energy excitations of the Kondo state can be considered as quasi-particles with an inverse effective mass ($1/m*$) and a residual interaction renormalized proportionally to $T_K$. These quasi-particles yield a narrow peak of width $T_K$ at the Fermi energy in the density of states called Kondo resonance.

\subsection{Linear transport of a quantum dot: Conductance and Shot Noise}
A quantum dot in the Kondo regime is thus a local Fermi liquid that we can precisely investigate. At equilibrium, transport is well described by the Landauer-Buttiker theory where the transmission of a channel $i$ is given by:
\begin{equation*}
T_i=\sin^2(\delta_i)
\end{equation*}
In the limit of low temperature and low frequency ($\hbar\omega\leq k_BT\ll eV$) the conductance and the symmetrized current noise are given by the formula~:
\begin{align*}
G=G_Q\sum_{i}T_i& &S_i=2eFI
\end{align*}
where the asymmetrized noise is defined \cite{Gavish2000}~as: 
\begin{equation*}
S^a_i(\nu)=\int^{+\infty}_{-\infty} d\tau <\delta I(t) \delta I(t+\tau)> e^{i 2 \pi \nu \tau}
\end{equation*}
and the symmetrized noise is~: 
\begin{equation*}
S_i(\nu)=S^a_i(\nu)+S^a_i(-\nu)
\end{equation*}
With our convention, for negative frequencies the asymmetric noise is related to emission processes, whereas absorption noise corresponds to positive frequencies. $F$ is called the Fano factor, which is given in the limit of non interacting quasiparticles by the relation~:
\begin{equation}
F=\frac{\sum_i T_i(1-Ti)}{\sum_i T_i}
\label{Fano_def}
\end{equation}
 
Actually, $F$ is a measurement of correlations and statistics of the charge carriers. In the tunnel limit, if all channels have  a low transmission, the noise is maximum and reaches the so-called Poissonian limit $F=1$, which corresponds to uncorrelated events. For non-interacting fermions, the formula \ref{Fano_def} is valid and always yields a subpoissonian noise: $F\leq 1$ with the limit $F=0$ for perfect transmissions. A super-poissonian noise $F\geq 1$ would be the signature of interactions or inelastic events.

For the different symmetries we obtain~:\\

\begin{tabular}{|l|c|c|c|}
\hline
symmetry &SU(2)&SU(4)& SU(4)\rule[-7pt]{0pt}{20pt}\\
\hline
filling&m=1&m=1 or 3&m=2\\
\hline
phase shift&$ \delta=\pi/2$  & $ \delta_1=\delta_2=\pi/4$  & $ \delta_1=\delta_2=\pi/2$\rule[-7pt]{0pt}{20pt}\\
\hline
transmission&T=1&$T_1=T_2=0.5$&$T_1=T_2=1$\rule[-7pt]{0pt}{20pt}\\
\hline
$G=G_Q\sum T_i$&$\frac{2e^2}{h}$ & $\frac{2e^2}{h}$  &$\frac{4e^2}{h}$\rule[-7pt]{0pt}{20pt} \\
\hline
$ F=\frac{\sum T_i(1-T_i)}{\sum T_i}$ &$F=0$ &$ F=0.5$ &$ F=0$\rule[-7pt]{0pt}{20pt}\\
\hline
\end{tabular}
\\

For SU(2), Kondo effect creates a resonant channel of width $~T_K$ at Fermi energy, which yields a perfect transmission through the dot and thus an absence of shot noise.

For SU(4), at quarter filling, transport takes place through two conducting channels of transmission $T_i=1/2$, which creates a strong partition noise. At this filling the electron-hole symmetry is broken, and the resonance of each channel can be seen as a global resonance, which is not centered at Fermi energy but shifted by $k_BT_K$.
At half filling (m=2), the electron-hole symmetry is recovered as well as the perfect conductance and the absence of partition noise.

An important point is that Fermi liquid theory can be pushed beyond the linear regime if the residual interaction between quasi-particles is taken into account. A quantum dot is thus an experimental test-bed for non-equilibrium Fermi liquid. In addition, experiments are expected reveal the universal behaviour of the Kondo state\footnote{it only depends on $T_K$ and not on the microscopic details of the device}.

\subsubsection{Asymmetry of the contacts}
If the contacts to the dot are not symmetric ($\Gamma_L\neq\Gamma_R$), transmissions have to be replaced by:
\begin{equation*}
T_i=\frac{4\Gamma_L\Gamma_R}{(\Gamma_L+\Gamma_R)^2}\sin^2(\delta)
\end{equation*}

The "bad" consequence for $SU(2)$ and half filling $SU(4)$ Kondo effect is that G is no longer unitary (the transmission is smaller than one) and partition shot noise occurs ($F\neq 0$), which hides the Kondo signature.

\subsection{Non-linear transport in the Kondo regime}
\subsubsection{Non-linear conductance}

Beyond equilibrium, quadratic terms in voltage, temperature and magnetic field appear in the conductance due to a residual interaction between quasiparticles of the Fermi liquid . However, Kondo effect is expected to exhibit universal behaviour even out of equilibrium. All response functions are governed by the Wilson ratio $R$, which is a measure of this residual interaction (see details in the appendix). In particular, in the strong coupling limit ($\frac{U}{\Gamma}\gg1$), the conductance becomes:

  \begin{equation}
  G(T,B,V_{sd})=G(0)\left[ 1-c_T\left(\frac{\pi T}{4T_K}\right)^2-c_B\left( \frac{\pi g\mu_BB}{4T_K}\right) ^2-c_V\left( \frac{\pi eV}{4T_K}\right) ^2\right] 
  \label{Gscaling}
  \end{equation}
   where the Fermi-liquid coefficients $c_T, c_B$ and $c_V$ depends only on the Wilson ratio $R$ through:
   \begin{equation}
   c_T=\pi^2\frac{1+2(R-1)^2}{3},\quad c_B=\frac{R^2}{4},\quad  c_V=\frac{1+5(R-1)^2}{4}.
   \label{Fermicoef}
   \end{equation}

 In our first experiments, we have used these coefficient to determine self-consistently $R$ and $T_K$ in a CNT dot in the SU(2) Kondo state (see also supplementary material in \cite{Ferrier2016}).

\subsubsection{Non-linear current and noise: effective charge}
Current and noise trough a quantum dot in the strong coupling ($T,V\ll T_K$ and $U/\Gamma\gg 1$ ) regime has been derived by different groups\cite{Sela2006,Gogolin2006,Mora2009,Sakano2011a} using different formalism.
For a symmetric dot ($\Gamma_L=\Gamma_R$), it appears that the current when developed until order $V^3$ (like in the previous section) can be recast like this:
\begin{equation}
I=G_QV-eP_{b_0}-eP_{b_1}-2eP_{b_2}
\label{ITheo}
\end{equation} 
where \begin{itemize} \item $P_{b_0}$ is the probability per unit of time that {\bf{one quasi-particle}} is elastically backscattered. This backscattering is not related to interaction effects but to the shift of the Kondo resonance when applying a voltage.
\item $P_{b_1}$ is the probability per unit of time that {\bf{one quasi-particle}} is backscattered because of the residual interaction.
\item $P_{b_2}$ is the probability per unit of time that {\bf{two quasi-particle}} are backscattered because of the residual interaction.
\end{itemize}

The shot noise for its part can be written:
\begin{equation}
S_I=2e^2P_{b_0}+2e^2P_{b_1}+2(2e)^2P_{b_2}
\label{SiTheo}
\end{equation}

These three probabilities are independent and very small compared to the main forward current $G_QV$.
The backscattering current ($I_b=G_QV-I$) can thus be seen as a current carried by two different charge carriers $e$ and $2e$, which create a Poissonian noise. The total current trough the dot can be thought as a main flow going forward from which sometimes bubbles of 1 or 2 particles are expelled as a counter-flow.
The ratio of the shot noise and the backscattered current is a direct measure of an average charge carrying the backscattered current.

Indeed we have the relation:
\begin{equation*}
\frac{S_I}{2I_b}=\frac{(e)^2P_{b_0}+(e)^2P_{b_1}+(2e)^2P_{b_2}}{[eP_{b_0}+eP_{b_1}+(2e)P_{b_2}]}=\frac{\left<q^2\right>}{\left<q\right>}
\end{equation*}

These coefficients in the general $SU(N)$ case at half filling are:
\begin{equation*}
P_{b_0}=\frac{N}{12}\frac{eV}{h}\left(\frac{eV}{\tilde{\Delta}}\right)^2
\end{equation*}
\begin{equation*}
P_{b_1}=\frac{N(N-1)(R_N-1)^2}{12}\frac{eV}{h}\left(\frac{eV}{\tilde{\Delta}}\right)^2
\end{equation*}
 \begin{equation*}
 P_{b_2}=\frac{N(N-1)(R_N-1)^2}{6}\frac{eV}{h}\left(\frac{eV}{\tilde{\Delta}}\right)^2
 \end{equation*}
 
 Where $\tilde{\Delta}$ is the width of the renormalized Kondo resonance.
 In the strong coupling limit ($U/\Gamma>>1$), Wilson ratio reaches the value $R_N=\frac{N}{N-1}$ and $\tilde{\Delta}=\frac{4T_K}{\pi}$. But the important point is that in the strong coupling limit {\bf{the effective charge does not depend on $T_K$}}.
 It can be expressed as:
 \begin{equation}
 e^*=\frac{P_1+4P_2}{P_1+2P_2}
 \end{equation}
 which makes $e^*$ a direct measurement of the two-particle scattering ($P_2$). For example it yields universal numbers $e^*/e=5/3$ for SU(2) and $e^*/e=3/2$ for SU(4) at half filling.

\section{Sample Fabrication and Current Noise Measurement}
The CNTs are first grown by chemical vapour deposition on an oxidized undoped silicon wafer \cite{kasumov07} and contacted with two contacts ($20\mathrm{~nm}$ thick palladium (Pd) for high frequency noise measurement and Pd($6\mathrm{~nm}$)/Al($70\mathrm{~nm}$) for low frequency noise measurement) separated by a distance of $400\mathrm{~nm}$. 

For high frequency measurement, the quantum detector (a superconductor / insulator / superconductor (SIS) junction) and the coupling circuit are designed and deposited in a single sequence, by angle evaporation of Al(70 nm)/AlOx/Al(100 nm). The sample is measured via filtered lines in a dilution refrigerator and cooled down to a temperature of $50\mathrm{~mK}$. The differential conductance is probed with a lock-in technique.

\subsection{Low frequency noise measurement}
To measure noise with high accuracy, our technique combines LC resonant circuit and a cryogenic amplifier based on high electron mobility transistor (HEMT) \cite{Arakawa2013,DiCarlo2006}.

\begin{figure}[htb] 
\centering 
\includegraphics[width=0.9\columnwidth]{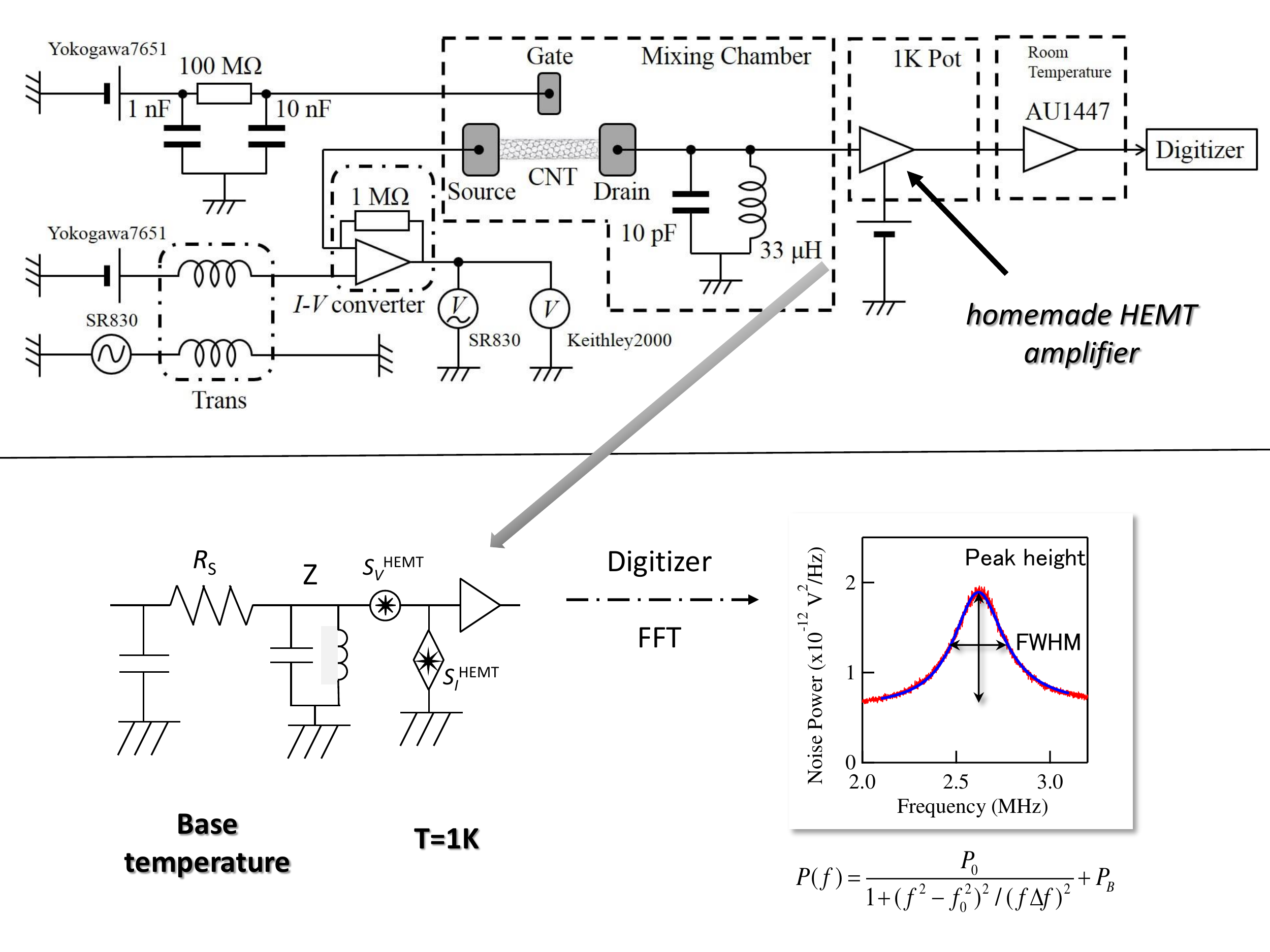}
\caption{Up) Measurement set-up used in the Osaka experiments. The HEMT based amplifier is detailed in \cite{Arakawa2013}. Bottom) Schematic circuit of the noise experiment used to extract the noise signal from the resonant peak. The resonant peak represents noise after amplification by a gain $A\approx10^6$ in power.}
\label{fig_setup} 
\end{figure}

The LC circuit acts as a bandpass filter at $f\approx 2.5\ $MHz before amplification by the cryogenic amplifier and the room temperature amplifier. The purpose of this set-up is to measure at frequency high enough to avoid 1/f noise contribution. Moreover the amplifier was made using 2 HEMT in parallel to reduce the 1/f noise of the HEMT themselves \cite{Arakawa2013}.

The basic principle of this technique is that the LC circuit converts current noise of the sample into voltage fluctuations through a narrow bandwidth around the resonance frequency. Then, the cryogenic amplifier transforms this voltage noise in a current noise through the $50\Omega$ output impedance. It realizes an impedance matching to transport this signal through a coaxial cable up to the room temperature amplifier. Finally this signal is recorded by a digitizer and a fast Fourier transformation is performed.

With this set-up, we obtained a sensitivity for the current noise power  below $10^{-29}\ A^2/Hz$ and measured an electronic temperature $T_e\leq 23\ $mK \cite{Hashisaka2009}.\\
\\

\textit{Noise measurement procedure:}\\
The Fourier transform of the voltage noise acquired after $LC$ circuit and amplifiers has a Lorentzian shape as shown in the bottom part of Fig. \ref{fig_setup}. The height $P_0$ and the width of this peak is extracted from a fit using the formula:
\begin{equation*}
S_V(f)=\frac{P_0}{1+\left(\frac{f^2-f_0^2}{f\Delta f}\right)^2}
\end{equation*} 
Then, the noise of the sample is extracted from a fit of the resonant signal.
\begin{equation}
P_0=A^2\left[\left(\frac{ZR}{Z+R}\right)^2S_I^{amp}+S_V^{amp}+\left(\frac{ZR}{Z+R}\right)^2S_I^{sample}\right]
\end{equation}

Parameters for the fit are: $Z$ the impedance of the resonant circuit at the resonance frequency, $A$ the gain of the amplifiers, $S_I^{amp}$ and $S_V^{amp}$ the noise of the amplifiers.

The impedance of the resonator is extracted from the width of the resonance: $\Delta f=(1/Z+G)/2\pi C$  with $C$ the capacitance of the LC resonator. $G=1/R$ is the conductance of the sample. If $G$ is tunable with a gate voltage, Z is extracted from the fit of the curve $\Delta f(G)$. Typically in this set-up, we have $Z\approx 100k \Omega\gg R_{sample}$.\\
This is an important parameter. This set-up is well suited to measure low impedance samples ($R\ll Z$) which ensures a good conversion between current noise of the sample to voltage noise at the input of the amplifier. From the equivalent circuit, neglecting noise of the amplifier itself, we have $S_V=\left(ZR/(Z+R)\right)^2 S_I$. For a given $R$, if $Z\gg R$ we obtain $S_V=R^2S_I$ which is the usual relation. However if $Z\ll R$ the sample is "short-circuited" by the resonator and we have $S_V=Z^2S_I$ which is very small since ($Z\ll R$).\\  

Other  parameters ($S_I^{amp},\ S_V^{amp}$ and $A$) are calibrated from the thermal noise measurement of the sample. Above $100\ $mK, we assume the electronic temperature is the same than the thermometer and use $S_I^{sample}=4k_BT/R$ to fit $P_0(R)$ between $100\ $mK and $650\ $mK. 

The typical values for the set-up  are $S_I^{amp}\approx 3 \, 10^{-28}$ $ A^2/Hz$ and $S_V^{amp}\approx 10^{-19}$ $V^2/Hz$ with a total gain $A=1.7 \, 10^6$.

\subsection{Measuring current noise in the quantum regime}
\label{sectionHFnoise}
\begin{figure}[h]
     \begin{center}
    \includegraphics[%
      width=0.65\linewidth,
      keepaspectratio]{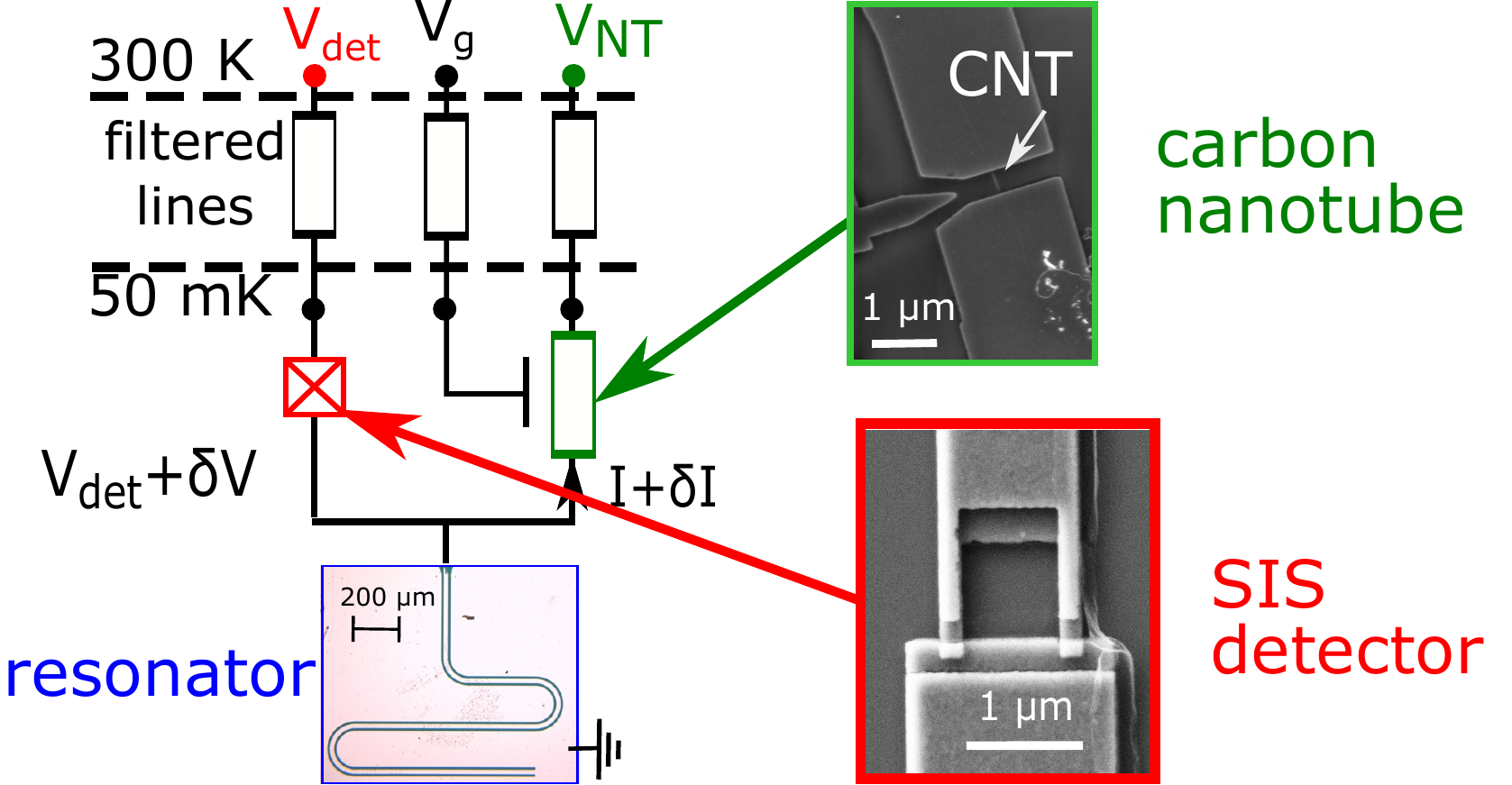}
    \end{center}
     \caption{Experimental setup. The CNT QD and the detector are connected together at one end of the central line of a coplanar waveguide of resonance frequencies $\nu_0=12$ and $\nu_1=31$ GHz. The SIS junction has a SQUID geometry such that its supercurrent can be suppressed by applying a small magnetic field. Figure adapted from \cite{Delagrange2018}.}
     \label{NoiseHF_setup}
\end{figure}

For high frequency noise measurement, the samples consist in CNT QDs, directly connected to coplanar waveguide resonators in a $\lambda/4$  configuration (fig.\ref{NoiseHF_setup}) such that the CNT can be dc-biased with a voltage $V_{NT}$. An electrostatic gate electrode is placed nearby the CNT, at a voltage $V_g$, allowing to tune the electrochemical potential inside the dot. A superconducting-insulator-superconducting (SIS) junction, used as a noise detector \cite{deblock2003,Billangeon2006,Basset2012a}, is also directly coupled to the resonant circuit. The signal emitted by the CNT is detected only at the resonance frequencies of the circuit, without any cut-off frequency up to the third harmonics \cite{Delagrange2018}. If the detector is biased below its superconducting gap $\Delta$ (i.e. if $e|V_D|<2\Delta$), photons of energy $h\nu>2\Delta-e|V_D|$ induce a photon-assisted tunnelling (PAT) current in the SIS junction. This DC current is proportional to the noise emitted by the CNT at frequencies $h\nu>2\Delta-e|V_D|$. Thanks to the frequency filtering by the resonator, a proper choice of $V_D$ allows to extract the noise at each circuit resonance \cite{Delagrange2018,Basset2012a}. 

\section{Low frequency shot noise}

\subsection{Conductance and Noise in the SU(2) Kondo state}

In a CNT quantum dot, spectroscopy of the states and detection of Kondo resonance are done by measuring the stability diagram $\frac{\partial I}{\partial V}(V_{sd},V_g)$. This is a 2D plot of the differential conductance as a function of source-drain voltage $V_{sd}$ and gate voltage $V_g$, which tunes the Fermi level of the dot.
 
\begin{figure}[h] 
   \centering 
   \includegraphics[width=0.8\columnwidth]{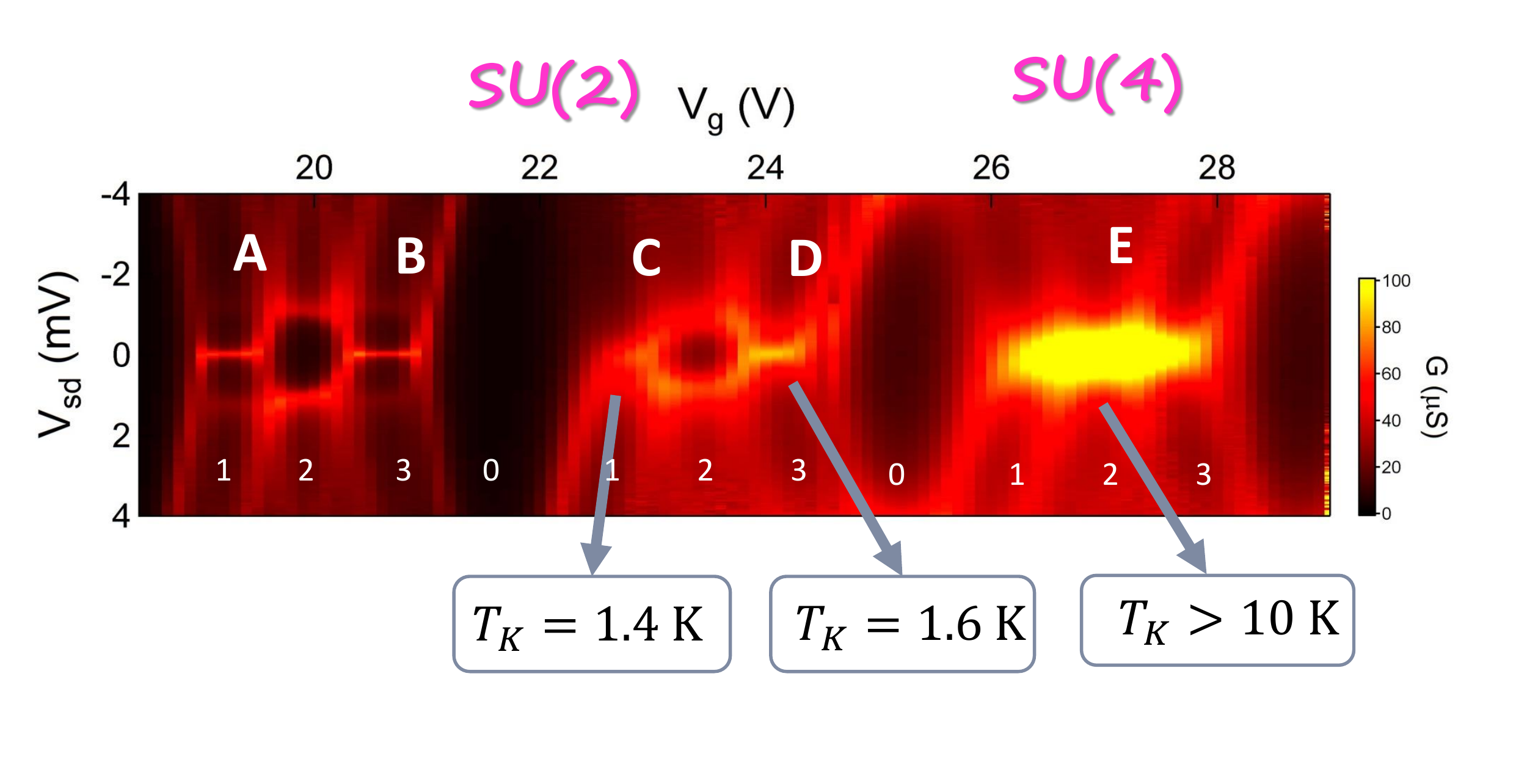}
   \caption{$SU(2)$ and $SU(4)$ Kondo ridges in the same CNT quantum dot. Stability diagram of the CNT dot measured at $T=20 $mK. Numbers indicate the filling factor. Experiments in the SU(2) Kondo state are done on ridge $D$ where $G\approx G_Q$. The SU(4) symmetry was investigated on ridge $E$ where $G\approx 2G_Q$ at half-filling ($N=2$).  
   }
   \label{3shell} 
   \end{figure}

 As the Kondo resonance provides a perfect transmission, the conductance is maximum at $V_{sd}=0$ whereas $G\sim 0$ in the Coulomb blockade regime. The Kondo state appears as a bright line at $V_{sd}=0$ parallel to $V_g$ axis called Kondo ridge in this image-plot.
 Depending on $V_g$ (see Fig.\ref{3shell}), we have observed $SU(2)$ (ridges A,B,C,D) as well as $SU(4)$ (ridge E) symmetry of the Kondo state. The first evidence is that, in the $SU(2)$ symmetry, Kondo ridges appears only for $N=1$ and $N=3$ whereas in the SU(4) symmetry we have detected the Kondo ridge for filling $N=1,\ 2$ and $3$.
 
 Indeed, since the dot-lead coupling $\Gamma$ depends on the gate voltage as well as the spin-orbit coupling and the valley scattering although to a lesser extent, it is often possible to observe both symmetry in the same CNT for different gate voltages \cite{Makarovski2007}.
 
 \subsubsection{Around equilibrium : a noiseless Kondo channel}
 
 In the SU(2) state  we could measure the residual interaction independently from two different methods. First , we have extracted the value of the Wilson ratio $R$ or equivalently interaction $U/\Gamma$ from the analysis of the Fermi coefficients of the non linear conductance. 
 We obtain $R=1.95$ and $T_K=1.6\ $K for the ridge $D$, which ensures that the strong coupling limit is achieved ($U/\Gamma\gg 1$) (see appendix).
 Then we did it directly from the value of the effective charge $e^*$ extracted from the non-linear noise. This constitutes a check of the non-equilibrium Fermi liquid theory and demonstrates the reliability of our noise measurements and analysis. 
 
\begin{figure}[h] 
  \centering 
  \includegraphics[width=0.95\columnwidth]{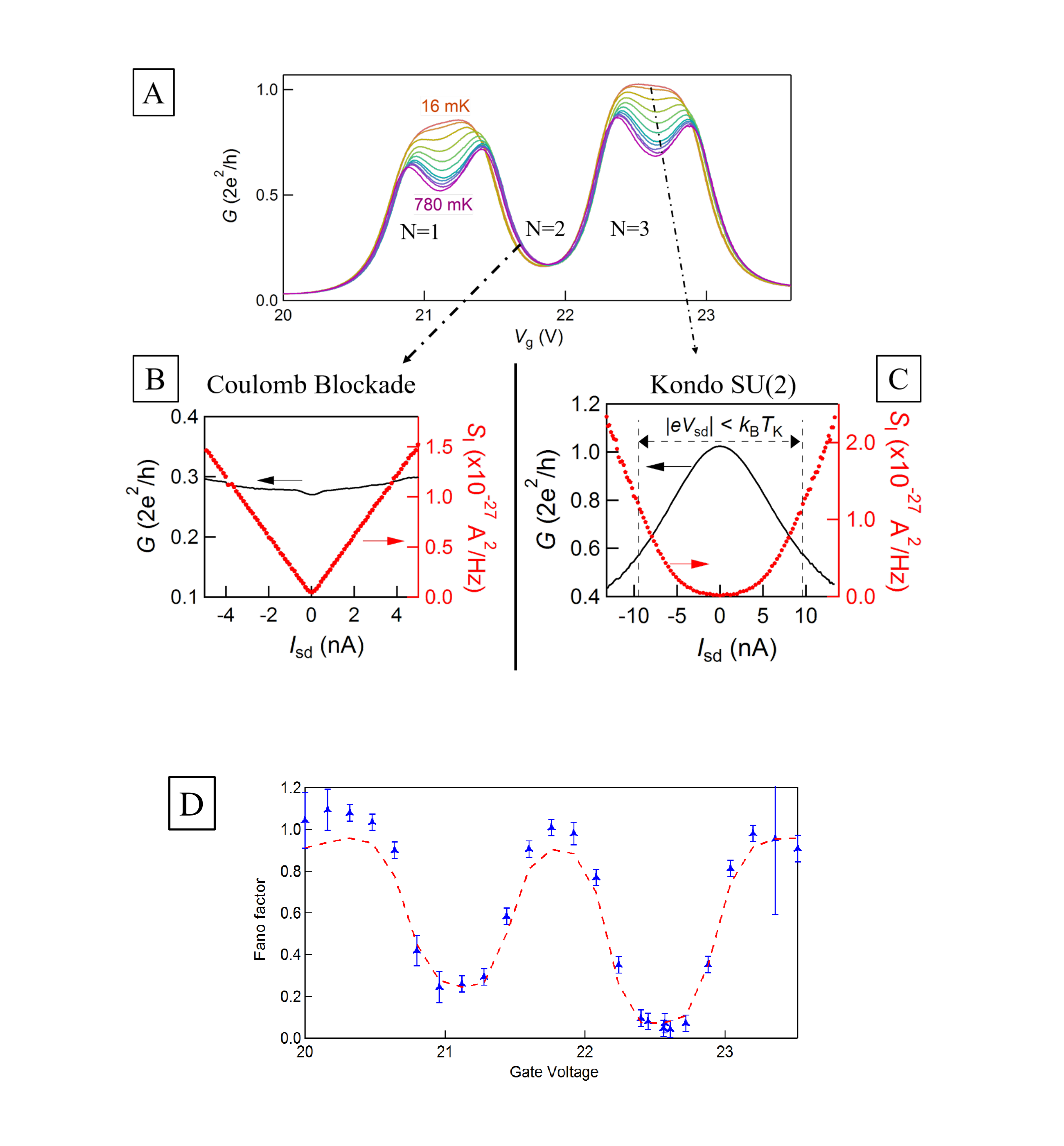}
  \caption{Conductance and noise in the SU(2) symmetry. Adapted from \cite{Ferrier2016}.  A) Conductance as a function of $V_g$ for ridges $C$ and $D$ on Fig \ref{3shell} at different temperatures. Kondo effect emerges for $N=1$ and $N=3$. The conductance reaches the unitary limit $G=G_Q$ on ridge $D$ (N=3). The Kondo effect progressively disappears when temperature increases ($G$ decreases at high temperature). B) $G$ and $S_i$ in the Coulomb blockade regime ($N=2$) as a function of current $I_{sd}$. Conductance is almost constant and low $G \approx 0.3 G_Q$ and the noise is linear  with a slope given by $F=1-T\approx 0.7$.   C) $G$ and $S_i$ in the Kondo regime ($N=3$) as a function of $I_{sd}$. $G$ is unitary ($T=1$) at $V_{sd}=0$ and strongly non-linear. Current noise is flat at $V_{sd}=0$ yielding a Fano factor $F\approx 0$ and then increases non-linearly at high current. The Kondo resonance thus provides a perfect transmission channel which produces no noise around equilibrium. \cite{Ferrier2016}. D) Fano factor as a function of $V_g$. $F$ is extracted from a linear fit of the noise $S_i$ as a function of $I_{sd}$. Red dotted line is the theoretical expectation $F=1-T$.
  }
  \label{SU2ridgeAndNoise} 
  \end{figure}

 The analysis of the conductance and noise in the SU(2) regime is shown in Fig. \ref{SU2ridgeAndNoise}. In the low voltage regime, since Fermi liquid picture holds, it is possible to analyse conductance and noise in the framework of Landauer-Buttiker theory\cite{Blanter2000,Martin1992} which is valid in the linear regime for non interacting quasi-particles. We thus have the relations:
 \begin{equation}
 G=G_Q\sum T_i\ \ \text{and}\ \ S_i=2eFI_{sd}\ \ \text{with}\ \ F=\frac{\sum T_i(1-T_i)}{\sum T_i}
 \label{Fano}
 \end{equation}
 $T_i$ is the transmission of channel $i$ and $F$ the Fano factor. This Fano factor has been extracted for all gate voltages by a linear fit at low current and displayed in Fig.\ref{SU2ridgeAndNoise}D.
 The results are perfectly consistent with transport through a single transmission channel as expected when the $K,\ K'$ degeneracy is lifted and $SU(2)$ Kondo effect shows up. Fano factor can be simply written $F=1-T$. 
 
 \underline{In the blockaded regime} ($N=2$), no Kondo resonance is detected, $G$ is very low and $F$ is close to $1$. In first approximation, in this elastic cotunneling regime,it has been shown\cite{Okazaki2013,Sukhorukov2001} that electronic transport can be seen as a Poissonian process where electrons are transmitted with a small probability and independently\footnote{This is only true in the elastic co-tunneling regime. At higher $V_{sd}$, correlations in the inelastic co-tunneling regime give rise to super-Poissonian noise.}. This explains perfectly the linear behaviour of the current noise shown on Fig. \ref{SU2ridgeAndNoise} B.
 
 \underline{On the Kondo ridge} the situation is opposite. At $V_{sd}=0$ the conductance is close to $G_Q$ and the Fano factor is almost $0$. Shot noise presents thus a plateau around $I_{sd}=0$ as can be seen on Fig. \ref{SU2ridgeAndNoise} C. In this regime, transmission is perfect and current flows without fluctuations\footnote{In our analysis the thermal noise is subtracted, which explains the zero noise at $I_{sd}=0$}.

 \subsubsection{Details on the extraction of the transmissions}
 Actually, to obtain a perfect agreement with theory in the linear regime we have to consider two possible transport channels. In the analysis presented above, we suppose only one channel participates since Kondo effect presents the SU(2) symmetry. The second channel (K') is shifted to higher energy due to the valley and spin-orbit splitting but it could participate to transport. To extract it, we can combine the value of $G$ and $F$ and calculate the values of the two transmission from Eq. \ref{Fano} as explained in details in the chapter on the crossover. We found a second channel with a transmission $T_2=0.05$ which does not depend on $V_g$. This channel doesn't play any role in the Kondo effect and explain that conductance is slightly higher than $G_Q$ ($G=1.02G_Q$) on the ridge $D$. This could be due to the "tail" of the energy dependence of the $K'$ channel. 
 The theoretical curve for $F$ on Fig. \ref{SU2ridgeAndNoise} takes it into account.
 
 \subsubsection{Non-linear noise}
 This is at higher voltage, in the non-linear regime that interactions  manifest themselves. In the SU(2) regime at half filling, Kondo effect creates $V^3$ corrections in current and noise (see Eq. \ref{ITheo} and \ref{SiTheo}). This explains the peculiar shape of shot noise in the Kondo regime displayed in Fig.\ref{SU2ridgeAndNoise}. Since in the unitary limit ($T=1$), the linear noise is $0$ ($F=0$) the first term is the $|V|^3$ contribution which gives this highly non-linear shape.
 
 The effective charge is extracted from the ratio of the non-linear parts of $S_i$ and $I_{sd}$. We have defined and computed the two quantities:
 \begin{equation*}
 I_K=I_{sd}-G(0)V_{sd}
 \end{equation*}
 \begin{equation*}
 S_K=S_i-2eF|I_{sd}|
 \end{equation*}
 which corresponds to $S_K=S_i$ in the ideal case ($T=1$). 
 The effective charge is then obtained by plotting $S_K(I_K)$ and fitting the linear part at low $I_{K}$ since $e^*$ is defined as $e^*=S_K/2|I_K|$.\\
 Note hat in the theory, the transmission is perfect ($T=1$). Thus the backscattered current $i_b=I_{sd}-G_QV$ and the shot noise do not contain any linear term in $V$ and $e^*$ is the ratio between these two non-linear quantities. However, if $T<1$ a linear contribution appears in backscattered current $i_b=(T-1)G_QV$ and in the shot noise $S_i=2eG_QT(1-T)V$, which is the usual partition shot noise. Since these two linear terms are not related to interaction effect, we extract experimentally $e^*$ from the ratio of the non-linear part of the current $I_K$ and the shot noise $S_K$.
 
 \begin{figure}[h] 
 \centering 
 \includegraphics[width=0.95\columnwidth]{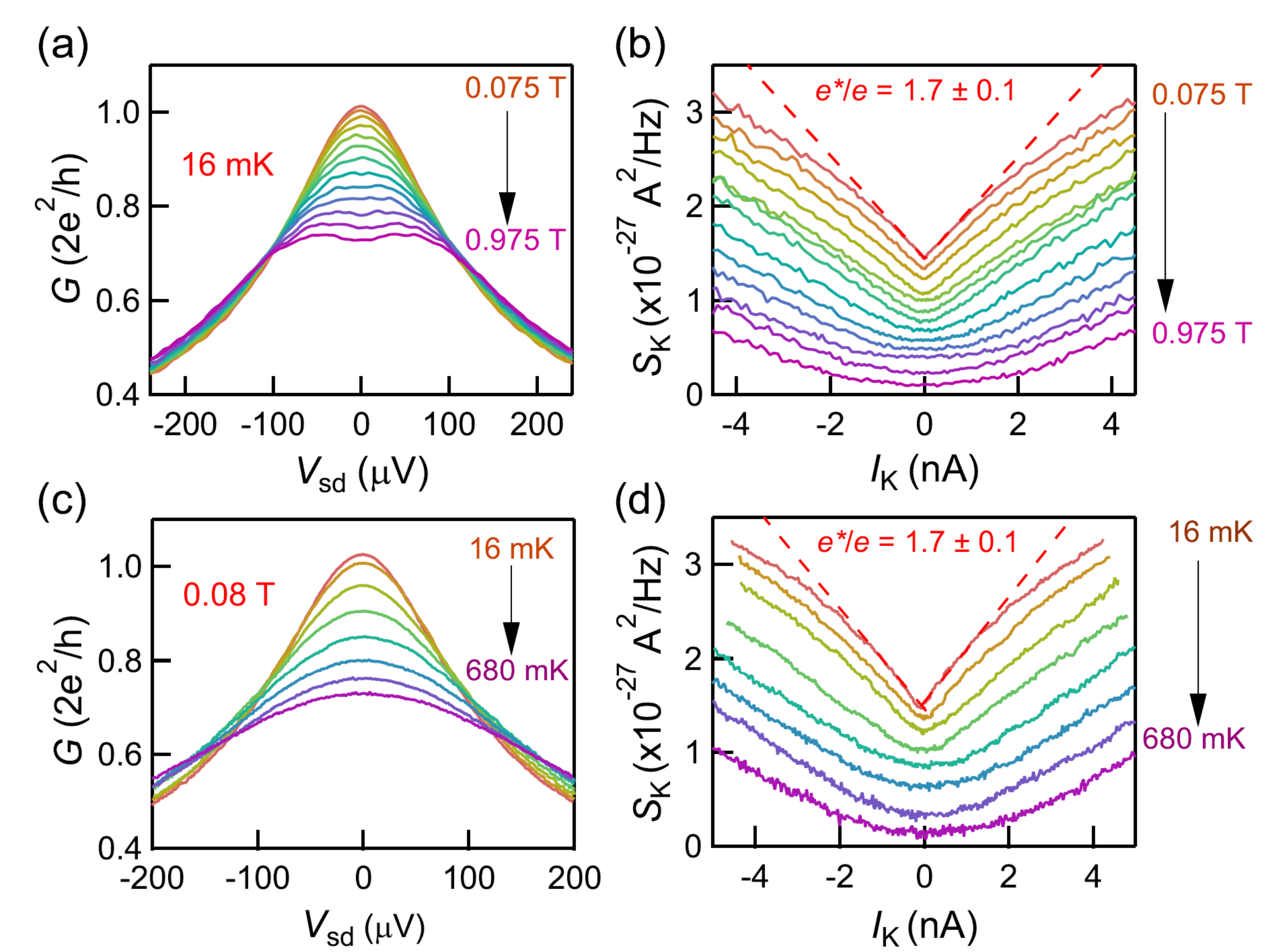}
 \caption{Evolution of the conductance and non-linear current noise with magnetic field (a and b) and temperature (c and d). Adapted from \cite{Ferrier2016}.}
 \label{FigEstardeTetB} 
 \end{figure}
 
 To check that $e^*$ is related to the Kondo resonance, we have also measured the field and temperature dependence of $e^*$. For the magnetic field evolution at $T=20\ $mK, we have kept the same definition with\footnote{The definition of $e^*$ at finite field is not so clear since we could define the backscattered current at each field as $I_K(B)=I-G(V=0,B)V$. }:
 \begin{equation*}
  I_K=I_{sd}-G(V=0,B=0)V_{sd}
 \end{equation*}
 
 At finite temperature, it is not possible to separate clearly thermal and shot noise when $G(V)$ is non-linear. However, based on theoretical proposal\cite{Fujii2010,Mora2009}, the effective charge can be defined from:
\begin{equation*}
S_K(T)=S_i-S_0(T)
\end{equation*} 
where $S_0(T)=2eFG(V_{sd}=0)V_{sd}\left(\coth(\frac{eV_{sd}}{2k_BT})-\frac{2k_BT}{eV_{sd}}\right)$.
The Fano factor $F$ and $G(V_{sd}=0)$ are evaluated at $16\ $mK.
For $I_K$, it is always the linear part evaluated at $16\ $mK which is subtracted : $I_K=G(V_{sd}=0)V_{sd}-I_{sd}$. Using the formula $S_K=2e^*|I_K|$, the effective charge is given by a linear fit at low current for all temperatures from $16\ $mK to $680\ $mK.
 The results plotted on Fig.\ref{FigEstardeTetB} show how $e^*$ vanishes with $T$ and $B$. Moreover, as expected in a Kondo system, the curves $e^*(B/T_K)$ and $e^*(T/T_K)$ can be rescaled on a single curve, which is a signature of the universality of the Kondo effect.

\subsection{Conductance and Noise in the SU(4) Kondo state}
\label{SU4section}
Now, we shift to the SU(4) Kondo state observed in ridge $E$ of Fig. \ref{3shell}. This ridge presents the typical behaviour expected in the SU(4) state with conductance $G=2G_Q$ maximum at half filling (N=2)

\subsubsection{Remark on the electron-hole symmetry in current and noise}
\begin{figure}[h] 
\centering 
\includegraphics[width=\columnwidth]{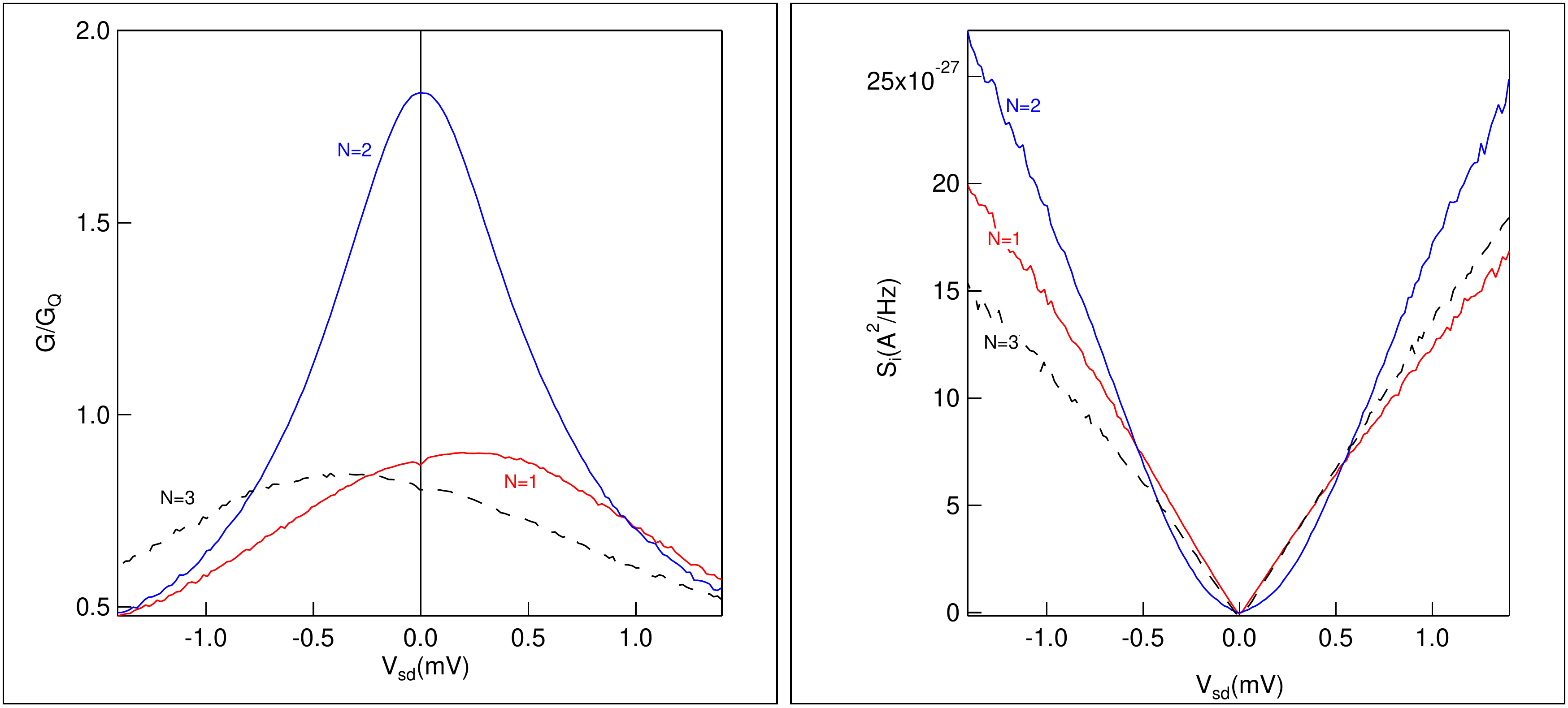}
\caption{Asymmetry in the conductance and noise in the $SU(4)$ state. The left/right asymmetry is $G_{N=2}/2G_Q\approx 0.92$. At half-filling (N=2) $G$ and $S_i$ are symmetric because of the electron-hole symmetry. For even number of electrons (N=1 and N=3) $G$ and $S_i$ become asymmetric with $V_{sd}$. The sign of the asymmetry change between $N=1$ and $N=3$ (1 hole). This can be written : $G_1(V_{sd})=G_3(-V_{sd})$ (same for $S_i$) as predicted in \cite{Mora2009}.}
\label{Asymetry} 
\end{figure}

Interestingly, electron-hole symmetry clearly appears in the current and shot noise measured through the CNT dot (see Fig.\ref{Asymetry}). At low bias ($eV_{sd}\ll T_K$), in the $SU(2)$ regime or in the $SU(4)$ regime at filling $N=2$, $G$ and $S_i$ are symmetric with $V_{sd}$ even if contact are not perfectly symmetric (left/right asymmetry). On the other hand, in the SU(4) regime at filling $N=1$ and $N=3$ (quarter filling), the electron-hole symmetry is broken and $G$ and $S_i$ become strongly asymmetric with $V_{sd}$. However, we can see from Fig.\ref{Asymetry} that we have the relation $G_1(V_{sd})=G_3(-V_{sd})$ (and the same for $S_i$) as predicted in \cite{Mora2009}. That means that transport properties are the same by changing $V_{sd}\rightarrow -V_{sd}$ provided that we also change filling from one electron to one hole $N=1\rightarrow N=3$.  In other words, a linear term appears in $G$ and $S_i$ which sign depends on the filling factor. The striking consequence for $G$ is that it is not maximum exactly at $V_{sd}=0$. This is because at quarter filling the Kondo resonance is shifted at energy $k_BT_K$ above the Fermi level \cite{Delattre2009}.
  
\subsubsection{Linear noise: signature of two channels transport}
\begin{figure}[h] 
\centering 
\includegraphics[width=0.9\columnwidth]{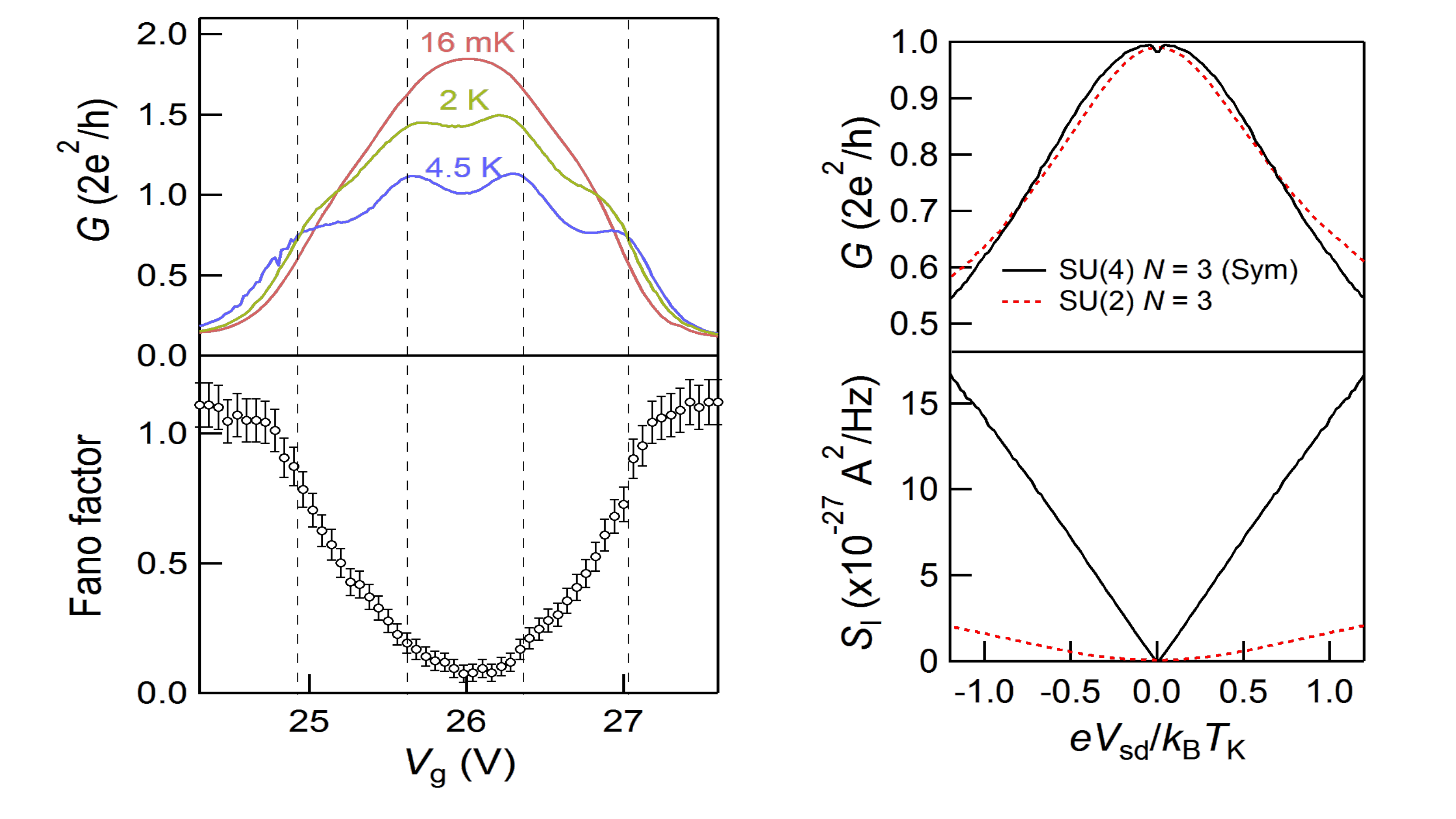}
\caption{Left) Conductance and Fano factor in the SU(4) Kondo state as a function of $V_g$. Right) Comparison of the conductance and shot noise in the SU(4) state at quarter filling (N=3) and the SU(2) state at half-filling (N=3). The absolute value of $G$ is the same and the shape versus the rescaled voltage $V_{sd}$ is similar. However the linear part of the noise is totally different. Fano factor is almost zero in the SU(2) case (one perfect channel) whereas Fano factor is close to $0.5$ in the SU(4) symmetry (two half-transmitted channels). Taken from \cite{Ferrier2016}.}
\label{LinearSU4} 
\end{figure}
      
Here also near $V_{sd}=0$, the Landauer-B\"uttiker formalism can be used. However, two channels have to be taken into account. We have extracted the Fano factor for all gate voltages from a linear fit of the current as in the previous case. This measurement (Fig. \ref{LinearSU4}) shows that two different Kondo ground states are formed at quarter filling ($m=1\ $ or $3$) and half filling ($m=2$). At quarter filling, we have measured $G=G_Q$ and $F=0.5$ which corresponds to two transport channels with $T_i=0.5$ as seen in section \ref{Fermiliquid}, which creates a strong partition noise. At half filling ($m=2$), conductance is $G=\frac{4e^2}{h}$ and we recover a "noiseless" Kondo resonance since current is carried by two perfect channels $T_i\approx 1$.
  
\subsubsection{Effective charge at filling N=2}
\begin{figure}[h] 
\centering 
\includegraphics[width=0.6\columnwidth]{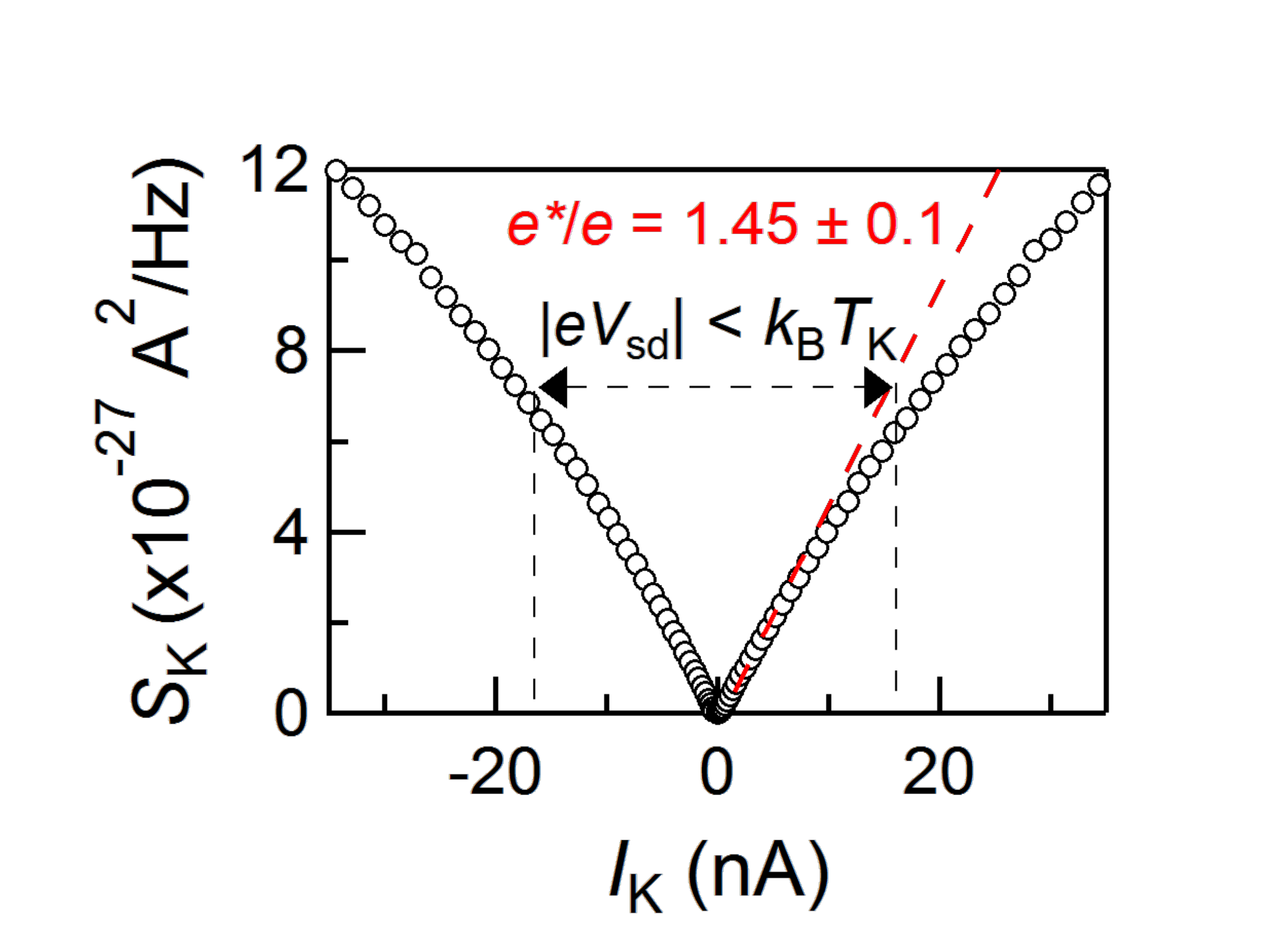}
\caption{Effective charge of the SU(4) Kondo state at half-filling. Taken from \cite{Ferrier2016}.}
\label{estarSU4} 
\end{figure}
          
We have extracted the effective charge $e^*$ at half filling. Indeed at quarter filling two difficulties make its extraction controversial. First, we have seen that a strong asymmetry appears in noise and current due to electron asymmetry. In this situation, how should we define $e^*$: only from the symmetric part of these quantities? On the other hand, we have also shown that the linear shot noise is very strong in this case since transport takes place trough two channels with $T_i=0.5$. The non linear noise, which has been detected in \cite{Delattre2009}, is thus a very small correction to this main linear contribution and is very difficult to extract accurately.
  
At filling $m=2$, we have computed the two quantities $I_K$ and $S_K$ to extract $e^*$ in the same way than for $SU(2)$. The result plotted on Fig \ref{estarSU4} shows that the relation is well linear at low current and the effective charge is $e^*/e=1.45\pm 0.1$. This result is also in very good agreement wit theory\cite{Mora2009,Sakano2011a} which predicts $e^*/e=3/2$.
The weakening of $e^*$, compared to $SU(2)$, corresponds to a decrease of the pair backscattering due to the weakening of quantum fluctuations of the spin on the dot in the SU(4) symmetry. 

\subsection{Crossover in field: effect of quantum fluctuations}
This part corresponds mainly to the results pusblished in \cite{FerrierPRL}.
The details on the effect of magnetic field on the CNT bandstructure is given in the appendix.

In a CNT, the orbital degeneracy can be lifted by a parallel magnetic field. By applying an in-plane field we have investigated in this same CNT the continuous crossover between $SU(4)$ and $SU(2)$ Kondo state at half filling ($m=2$). This has allowed us to investigate directly how the symmetry group and more precisely the associated quantum fluctuations control the Kondo state \cite{FerrierPRL}.
The main point, making the originality of our experiment, is that we pass continuously from $SU(4)$ to $SU(2)$ at constant filling factor (gate voltage) without killing the Kondo state (as it is the case when we move the gate voltage and change the filling factor) as explained in Fig. \ref{crossover_scheme}. 

As explained in the appendix, quantum fluctuations of the total spin depends on the symmetry group $SU(N)$. These fluctuations and thus the residual interaction are quantified by the Wilson ratio: $R=1+\frac{1}{N-1}$ when the symmetry is well defined. By changing continuously the symmetry of the  Kondo state from $SU(4)$ to $SU(2)$ at constant half-filling we could thus investigate the evolution of transport properties in the Kondo regime when quantum fluctuations continuously increase.

\begin{figure}[h] 
\includegraphics[width=0.9\columnwidth]{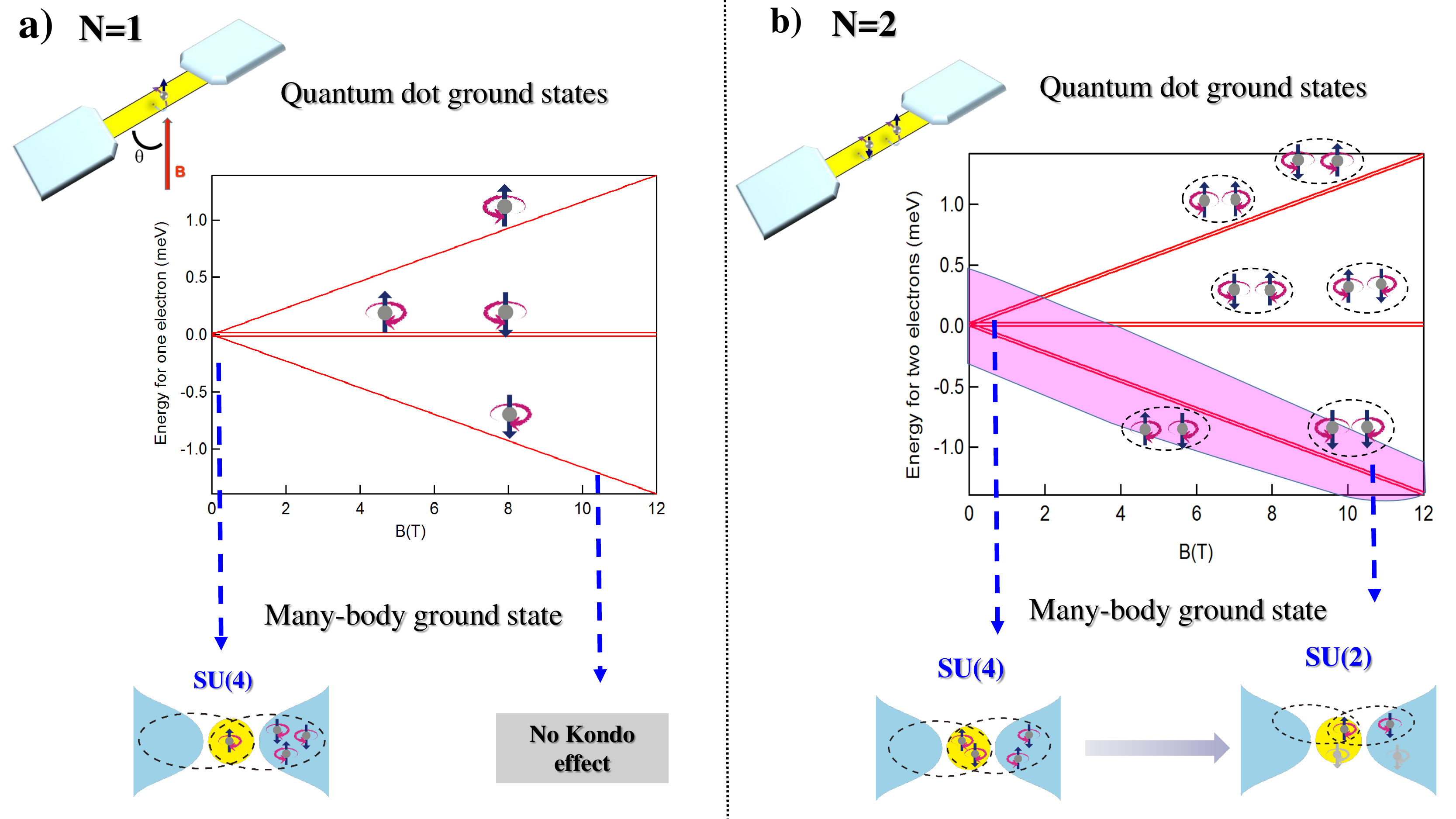}
\caption{Single particle energy spectrum of the CNT dot for the filling factors $N=1$ and $N=2$ (half filling). The angle of the field is such that $g_{orb}\cos\theta\approx \frac12 g_S=1$. a) For $N=1$, at B=0, the four degenerate states can form a $SU(4)$ Kondo singlet. It disappears at finite field since the degeneracy is lifted. b) For $N=2$, each line of the spectrum is doubly degenerate (see text). At B=0 the six degenerate states form the $SU(4)$ Kondo state. At finite field, it continuously evolves to an $SU(2)$ Kondo state since only two degenerate states form the ground state. The complete ground states are depicted in Fig. \ref{SU4GS}, in the appendix and in \cite{FerrierPRL}.}
\label{crossover_scheme} 
\end{figure}

In our experiment, the angle $\theta$ between the field and the axis of the tube is such that $g_{orb}\cos\theta\approx \frac12 g_S=1$,where $g_S$ and $g_{orb}$ are the spin and orbital Land{\'e} factors (see appendix).

As illustrated on Fig.\ref{crossover_scheme}, for a CNT, magnetic field acts on the spin with the Zeeman energy $E^{spin}=\frac12\sigma g_S\mu_BB$ whereas only the parallel component $B_{\parallel}=B\cos\theta$ acts on the orbital momentum adding a term $E^{orb}=\tau g_{orb}\mu_BB\cos\theta$, where  $\sigma=\pm1$ refers to the spin direction and $\tau=\pm1$ to the valley quantum number. Since $g_{orb}\cos\theta\approx \frac12g_S=1$, both spin and valley degrees of freedom are shifting with the same energy $\mu_BB$. Consequently the total magnetic energy for one electron is just $E=(\sigma + \tau)\mu_BB$ (see Fig.\ref{crossover_scheme}a). 

If the dot is filled by two electrons with two quantum numbers, there are $6$ degenerate states described in the Fig. \ref{crossover_scheme}. When the field is applied, since spin and orbital energy are equal, states remain degenerate two by two. Indeed, as plotted on Fig. \ref{crossover_scheme}b, a straightforward calculation shows that two states are shifted by $\Delta E=-2\mu_BB$, two are not shifted and the last two are shifted by $\Delta E=+2\mu_BB$. As a consequence, the ground state remains doubly degenerate when applying the magnetic field. At $B=0$ the six degenerate states contribute to the Kondo effect giving the $SU(4)$ symmetry. At high field, when only two states are degenerate within the characteristic energy scale of the Kondo effect $k_BT_K$ the usual $SU(2)$ Kondo state is recovered. 
 
\subsubsection{Conductance Stability Diagram in the Crossover}
\begin{figure}[h] 
\includegraphics[width=0.9\columnwidth]{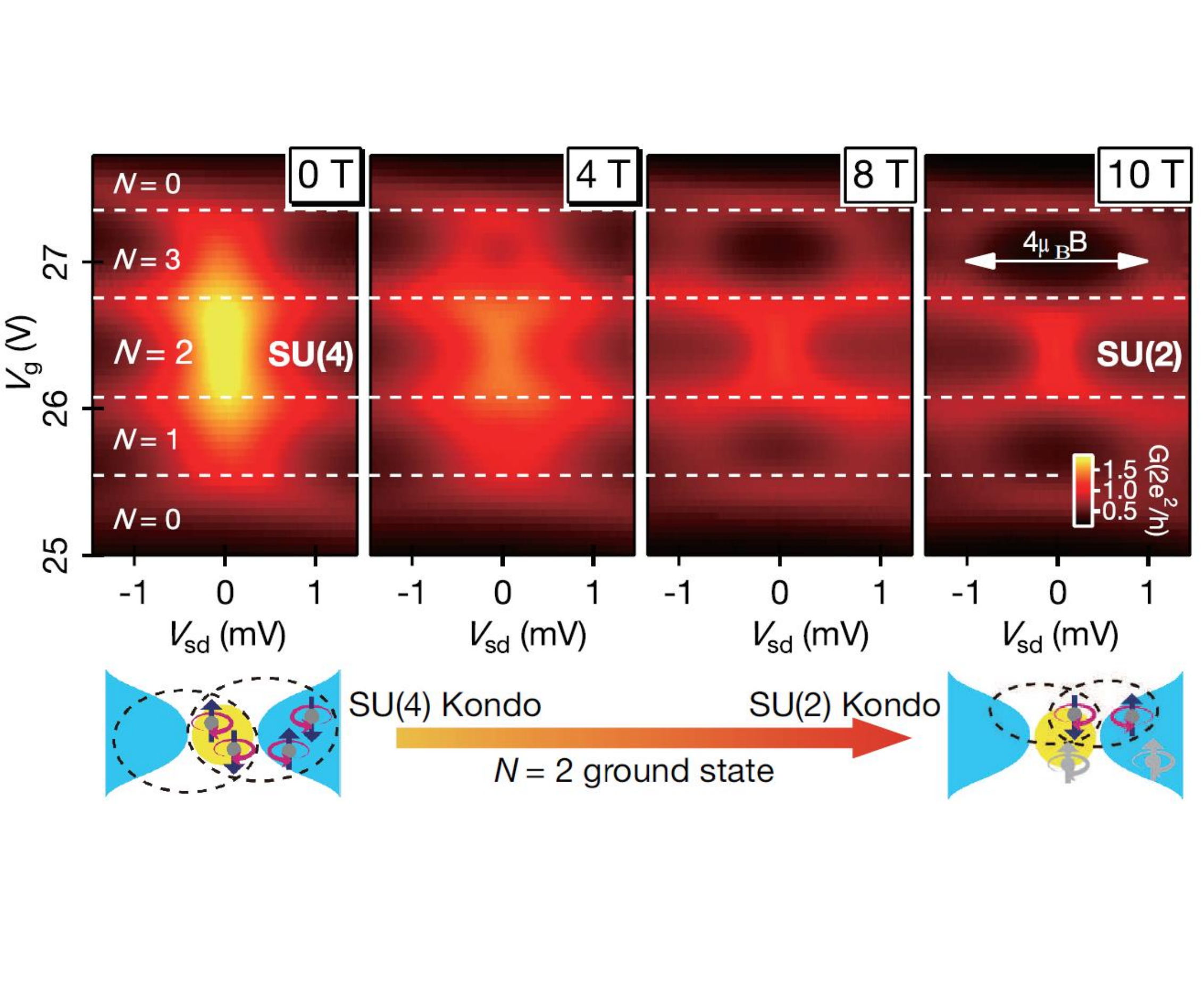}
\caption{Stability diagram along the crossover between the $SU(4)$ and the $SU(2)$ symmetry. Contour plot of $G$ as a function of $V_g$ and $V_{sd}$ at $T=16\ $mK. The Kondo resonance produces the bright vertical line at $V_{sd}=0$. This ridge disappears at high field for $m=1$ and $m=3$. It is split into two satellite peaks at high $V_{sd}$ separated by $e\Delta V_{sd}\approx 4\mu_BB$. At $m=2$, $G$ decreases but it remains maximum at $V_{sd}=0$. Below is a sketch of the evolution of the many-body ground state at $m=2$. Taken from \cite{FerrierPRL}.}
\label{Fig2cross} 
\end{figure}
 
Here also, the Kondo state has been characterized by measuring the stability diagram of the CNT quantum dot at different fields up to $12\ $T. At $B=0.08\ $T (to suppress superconductivity of Aluminum contact), this is the same SU(4) Kondo state as in section \ref{SU4section}. The result is displayed on Fig.\ref{Fig2cross}. At low field, the Kondo resonance is detected for all the fillings $m=1,2\ $ and $3$. When increasing the field, resonances at quarter filling $m=1$ and $m=3$ progressively disappear whereas it remains at half filling ($m=2$) until $B=12\ $T. On the contour plot, we can see the bright vertical lines for $N=1$ and $N=3$ become dark denoting that the resonance split and $G$ becomes minimum at $V_{sd}=0$. Two satellites appear at finite $V_{sd}$ following a slope close to $eV_{sd}=2\mu_BB$ as expected from the simplified picture of Fig. \ref{crossover_scheme}a for the single particle spectrum. On the other hand, for $N=2$, the Kondo ridge remains until $B=12\ $T, although its intensity decreases. At $B=0$, as seen in section \ref{SU4section}, $G=1.85G_Q$ which is almost the expected value in the unitary $SU(4)$ limit ($2G_Q$). At high field, it decreases to $G\approx G_Q$ as expected in the $SU(2)$ limit. 
 
\subsubsection{Linear Noise in the Crossover}

\begin{figure}[h] 
\centering 
\includegraphics[width=0.6\columnwidth]{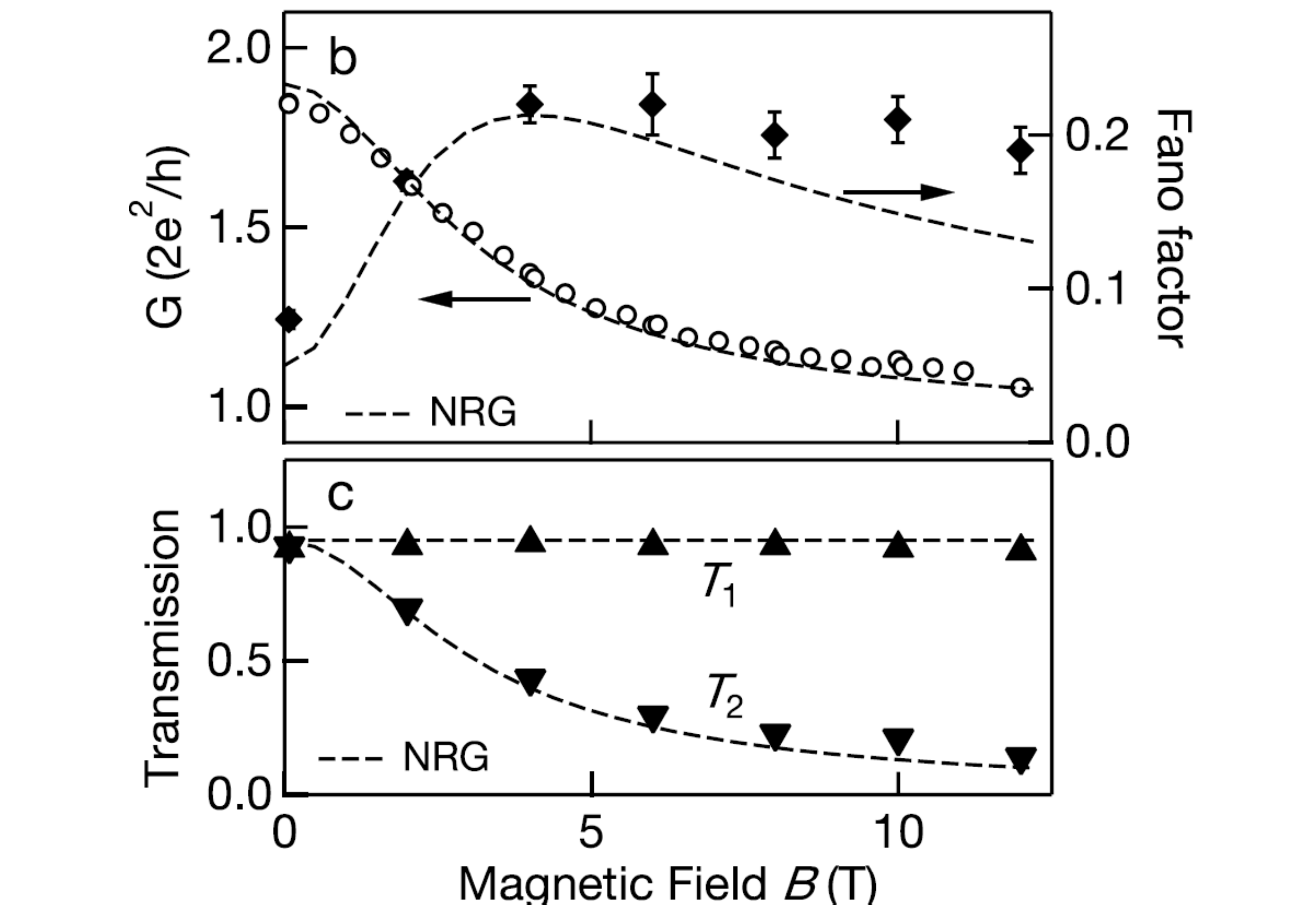}
\caption{Evolution of the linear transport along the $SU(4)$/$SU(2)$ crossover. Symbols represent experimental results. Dotted lines are obtained from NRG computation (see section \ref{NRGcrossover}. Upper part) $G$ and Fano factor as a function of magnetic field. Conductance decreases from $1.8G_Q$ in the $SU(4)$ state to $G_Q$ in the $SU(2)$ state. F is obtained from a linear fit of the current noise as a function of $I_{sd}$ at each field. Bottom) Evolution of the two transmissions $T_1$ and $T_2$ as a function of magnetic field. Transmissions are extracted from $G$ and $F$ using Eq.\ref{Fano}. One channel remains perfectly open during the crossover whereas the other one progressively disappears. Taken from \cite{FerrierPRL}.}
\label{Fanocrossover} 
\end{figure}
 
 This interpretation is independently confirmed by the noise measurement. The evidence of this crossover is the change of the number of transmission channels. At half filling, we have seen that transport occurs through two perfect channels in the $SU(4)$ symmetry whereas only one perfect single channel contributes in the $SU(2)$ symmetry.
 The evolution of transmission during the crossover can be directly extracted from conductance and noise measurement. Indeed, using Eq.\ref{Fano} for two channels we obtain $T_{1,2}=g/2\pm 1/2\sqrt{2g(1-F)-g^2}$. The results displayed in Fig.\ref{Fanocrossover} clearly show that one transmission $T_1$ remains constant and perfect along the crossover whereas the second one $T_2$ vanishes. At the same time, the conductance drops from $G=1.85G_Q$ to $G=G_Q$. This demonstrates that the dot evolves from the $SU(4)$ Kondo state with two perfect channels to the $SU(2)$ symmetry with a single perfect channel.
 
\subsubsection{Comparison with NRG computation}
\label{NRGcrossover}

\begin{figure}[h] 
\includegraphics[width=0.9\columnwidth]{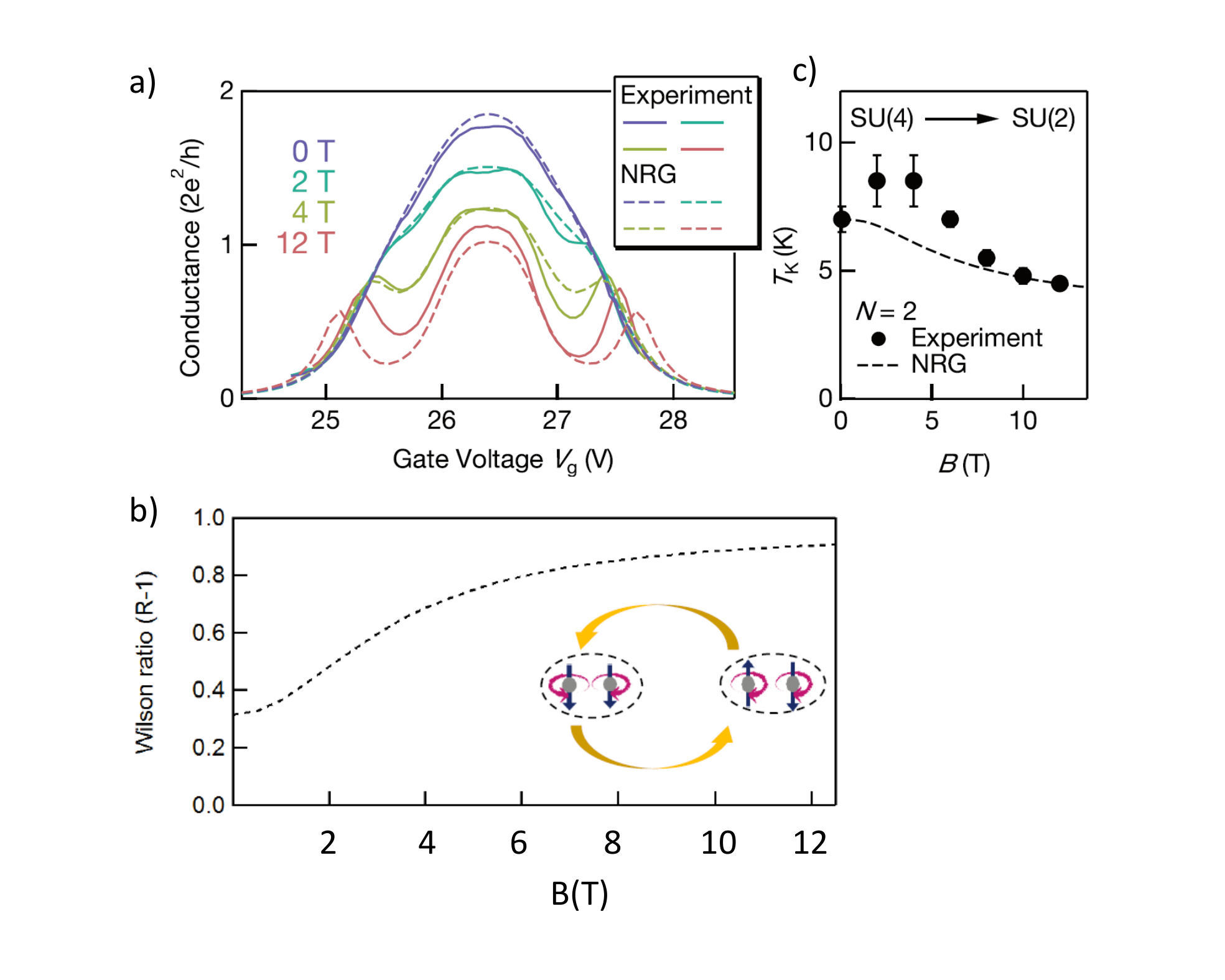}
\caption{Comparison with NRG computation. Extraction of Wilson ratio. a) Comparison of zero bias conductance between experiment and NRG calculation (dotted lines) foe several magnetic fields. This yields the wilson ratio for each value of $G(V_g)$. b) Wilson ratio as a function of $B$. $R$ is computed by NRG calculation with parameters yielding the successful comparison with experiment for $G$ and $F$ at $N=2$ for every magnetic field. c) Field dependence of $T_K$ at $N=2$. The dashed line is the result of the NRG calculation multiplied by a global pre-factor $1.3$ to fit the experimental value at $B=0$. The NRG $T_K$ is computed from the Wilson ratio $R$ at each field. Details for the experimental $T_K$ are given in the appendix. Figure adapted from \cite{FerrierPRL}.}
\label{Fig_NRG_wilson} 
\end{figure}

In addition, these results and interpretations are strongly supported by a very good agreement with NRG calculations, which confirms that Kondo regime is well established ($\frac{U}{\Gamma}\approx 3.15$) as well as the orientation of the field and the fact that $\Delta_{so}$, $\Delta_{KK'}$ and exchange energy $J_{KK'}$ can be neglected.
The details are given in ~\cite{Teratani2016} and in the appendix. 
Here we present the simplest case where $\Delta_{so}$, $\Delta_{KK'}$ and exchange energy $J_{KK'}$ are set to zero. We have successfully reproduced the complete shape of the zero-bias conductance $G$  as a function of $V_g$ as displayed in Fig.\ref{Fig_NRG_wilson}a. The solid lines are experimental results for $B=0$, 2, 4, and 12~T, whereas the dashed lines correspond to the NRG calculation for the same magnetic fields using the parameters $U/\Gamma=3.15$ ($U$ is the charging energy,  $\Gamma$ is the coupling strength) and $g_{\rm orb}\cos\theta=1$. This successful comparison allowed us to compute $T_K$ and the Wilson ratio $R$~\cite{Wilson1975}  from the NRG parameters as shown in Fig.\ref{Fig_NRG_wilson}. (see appendix for the definition of the present Wilson ratio in the multichannel case).  Thus, Fig. \ref{Fig_NRG_wilson}c shows how quantum fluctuations continuously increase along the crossover when the magnetic field increases. The computed  $T_K$ is in good agreement with experimental values in the two limiting cases $SU(4)$ and $SU(2)$ as shown in Fig.\ref{Fig_NRG_wilson}b.  The decrease of $T_K$ in the $SU(2)$ state denotes an enhancement of the lifetime ($\frac{h}{k_BT_K}$) of the Kondo resonance related to the increase of the fluctuations.
 
 \subsubsection{Effective charge in the crossover}
 \begin{figure}[h] 
 \centering 
 \includegraphics[width=0.7\columnwidth]{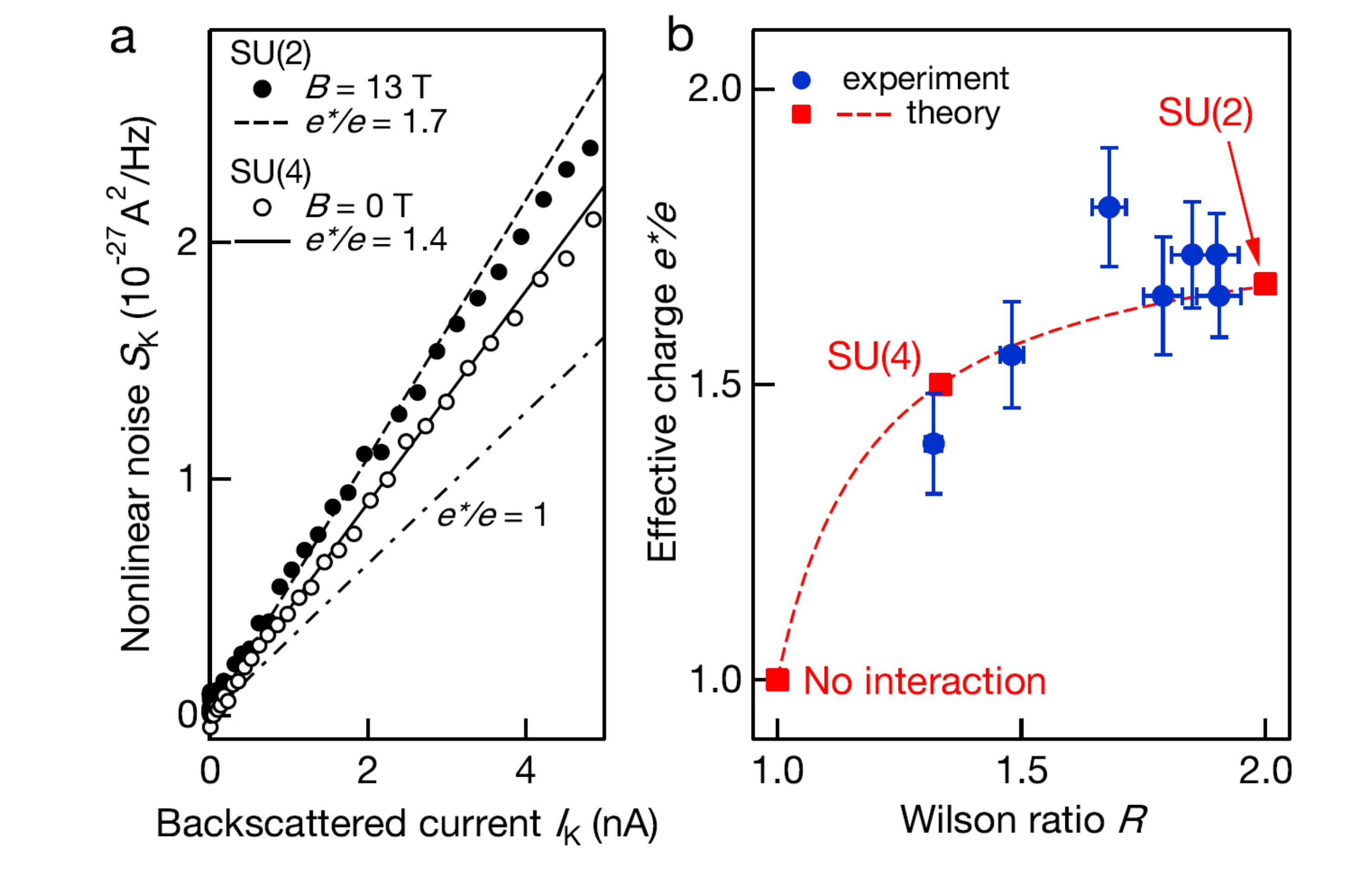}
 \caption{Evolution of the effective charge along the crossover between the $SU(4)$ and the $SU(2)$ symmetry. a) Non-linear noise as a function of backscattered current $I_K$ at $B=0$ (SU(4)) and $B=13\ $T (SU(2)). The solid and dashed lines are results of the linear fits yielding $e^*/e=1.4\pm 0.1$ for SU(4) and $e^*/e=1.7\pm 0.1$ for SU(2). b)Filled circles represent experimental $e^*$ as a function of $R$, extracted from NRG fit of conductance at the same field displayed in Fig. \ref{Fig_NRG_wilson}. $e^*(B)$ is shown in supplementary of ref \cite{FerrierPRL}.  Square symbols are theoretical predictions for SU(4), SU(2) and non interacting electrons. Dashed line is the extended theoretical prediction given in Eq. \ref{effective_charge_wilson}. Taken from \cite{FerrierPRL}.}
\label{Estarcross} 
\end{figure}

We have used the same procedure explained in the previous sections to extract $e^*$ for each magnetic field. 
At $T=0$, $e^*$ is experimentally defined as $S_K=2e^*|I_K|$. $S_K$ is the non-linear part of the noise $S_K=S_I-2eFI_{sd}$ and $I_K$ the non-linear part of the current $I_K=I_{sd}-2G(0)V_{sd}$.
 
Figure \ref{Estarcross} represents $S_K$ as a function of $I_K$ at $B=0.08\ T$ ($SU(4)$ state) and $B=13\ $T ($SU(2)$ state). The dotted lines show the result of the linear fit which yields $e^*/e=1.4\pm 0.1$ for $SU(4)$ and $e^*/e=1.7\pm 0.1$ for $SU(2)$ in perfect agreement with theory in these two limiting cases. We found that $e^*$ almost continuously increases as $B$ increases from 0 to 13~T. $e^*(B)$ is displayed in the supplementary material of ref \cite{FerrierPRL}. Knowing $R$ as a function of $B$ [Fig. \ref{Fig_NRG_wilson}], we represent $e^*$ as a function of $R$ in Fig. \ref{Estarcross}. Clearly, the effective charge gradually increases as $R$ increases. This graph illustrates how quantum fluctuations (namely, $R$) affect the two particle scattering ($e^*$). It demonstrates that $e^*$ is a relevant experimental measure to quantify quantum fluctuations.
  
Interestingly, this result can be reproduced if we push the theory beyond its domain of validity. For a well defined symmetry $SU(N)$, theory predicts that the Wilson ratio and the effective charge in the Kondo region are given only by $N$ such that $R =1+ \frac{1}{N-1}$ and $\frac{e^*}{e} = \frac{N+8}{N+4}$, respectively~\cite{Sakano2011}. We extend this relation in the broken symmetry region, yielding~:
\begin{equation}
  \frac{e^*}{e}=\frac{1+9(R-1)}{1+5(R-1)},
  \label{effective_charge_wilson}
\end{equation}
which is superposed as a dashed line in Fig.~\ref{Estarcross}(b). Our experimental measurement is well reproduced by this relation continuously along the crossover from $SU(4)$ to $SU(2)$ even in the intermediate symmetry region. This emphasizes how nonequilibrium properties ($e^*$) and equilibrium quantities ($R$) are intricately linked in quantum many-body states.
 
\subsection{Last remark on the symmetry groups: SU(4) vs SU(2)$\times$SU(2)}
Our result also shows that Kondo state at B=0 is really a $SU(4)$ state and not $SU(2)\otimes SU(2)$.
In a CNT quantum dot at half filling ($N=2$) at $B=0$, the $SU(4)$ symmetry of the Kondo resonance could be broken into a $SU(2)\otimes SU(2)$ symmetry~\cite{Sakano2012,Nishikawa2012} if the inter-valley Coulomb interaction is smaller than the intra-valley one and the difference is larger than $k_BT_K$. In this situation, the spin-$SU(2)$ Kondo state emerges almost independently in $K$ and $K'$ valleys yielding two separate $SU(2)$ conducting channels in parallel. Consequently, the conductance reaches $\frac{4e^2}{h}$ with two perfect channels as in a $SU(4)$ state. Nevertheless, the $SU(4)$ and the $SU(2)\otimes SU(2)$ states can be distinguished by the conductance and the non-linear noise measurement as follows. 
 
In the $SU(2)\otimes SU(2)$ symmetry, the two independent channels yield the same effective charge as the $SU(2)$ symmetry ($e^*/e=5/3$) whereas it is smaller in the $SU(4)$ state ($e^*/e=3/2$).
Our observation of $e^*/e \simeq 3/2$ at $B=0$ is already a first evidence that the Kondo state possesses the SU(4) symmetry. Although a contact asymmetry or a small value of $\frac{U}{\Gamma}$ could be responsible for a smaller $e^*$ than the expected $SU(2)$ value, the value of the measured conductance $G=\frac{3.7e^2}{h}$ already confirms that both contact symmetry and $\frac{U}{\Gamma}$ are high enough to prevent any suppression of $e^*$. 
 
Most importantly,  the fact that we detect an enhancement of $e^*$ in the crossover is a clear evidence that the symmetry at $B=0$ is no doubt $SU(4)$ since in the $SU(2)$ limit we obtain the expected value of $5/3$ which proves that it is not reduced by the contact asymmetry or weak coupling. This statement is also supported by the large value of $T_K$ in the $SU(4)$ symmetry which also prevents this symmetry breaking. 
 
\subsection{Comparison with previous experiments}
Before measurements presented in this paper, $3$ different experiments have been carried out and report a Kondo contribution to the non-linear noise \cite{Delattre2009,Yamauchi2011,Zarchin2008}. 
 
The first point to emphasize is that the noise in the linear regime had not been measured previously. To observe the linear regime we need to achieve $k_BT\ll eV\ll k_BT_K$. In the 2DEG experiments $T_K$ was too small ($700$ and $300\ $mK) to enter this regime whereas the CNT experiment of Delattre \textit{et al}\cite{Delattre2009} was performed at too high temperature. 
 
Another important point is to perform the experiment in the unitary limit of the Kondo effect. This requires left/right symmetry of the contacts and to be in the Kondo limit $\frac{U}{\Gamma}\gg 1$. The Kondo limit is important to ensure universality of the effective charge. Otherwise its value depends on $U$. 
Left/right asymmetry of the dot can drastically changes the result as calculated by Mora \textit{et al} \cite{Mora2009}. Moreover asymmetry yields also a strong partition noise which has to be subtracted to obtain the non-linear noise and makes the uncertainty on the experimental result rapidly increasing. It is thus difficult to perform a good quantitative measurement.
These two points were not reached in the 2DEG experiments. In both cases conductance does not reach $2G_Q$ and more importantly, the Kondo ridge is not flat (as a function of $V_g$) suggesting that the dot is not in the universal Kondo limit.
  
Finally, the experiment on the CNT performed in \cite{Delattre2009} was in the SU(4) Kondo state at filling $N=1$ whereas in this work we have analysed the non-linear noise at half filling ($N=2$).
 
\subsection{Conclusion for the Low Frequency Noise}

Using a carbon nanotube as a quantum dot, we have obtained a Kondo state which could be changed from the $SU(2)$ to the $SU(4)$ symmetry while remaining in the unitary limit. We have achieved a perfect agreement with theory, which provides a precise understanding of the link between non-linear noise, Kondo correlations and symmetry of the ground state.
 
First, we have demonstrated that around equilibrium ($eV\ll T_K$), the delocalized state transmits perfectly the current through the dot, without any backscattering. The quantum dot is thus totally silent.
Then, we have shown that interaction emerges beyond equilibrium at larger voltages. Our work shows that they can be characterized by an effective average charge $e^*$, extracted from the non-linear part of the shot noise. Our measurement that $e^*=5/3$ and $e^*=3/2$, at half filling in the $SU(2)$ and $SU(4)$ symmetry respectively, demonstrates the appearance of backscattered electrons pairs when a current flows through a Kondo impurity.
Finally, since a Kondo state can be described as a local Fermi liquid, our result can be seen as an experimental confirmation of the extension in the non-equilibrium regime of the Fermi-liquid theory.
     
\section{High frequency noise mesurements}

We now present the results obtained on the measurement of the emission noise in the quantum regime ($h \nu \gg k_BT$) on carbon nanotubes quantum dots in the Kondo regime with $SU(2)$ or $SU(4)$ symmetry. The noise detection is realized by using an on-chip quantum detector coupled to the CNT as described in section \ref{sectionHFnoise}. This setup allows also to measure the DC transport properties of the CNT quantum dot. 

\subsection{Conductance in the $SU(2)$ Kondo regime}
\begin{figure}[htb]
     \begin{center}
    \includegraphics[%
      width=0.65\linewidth,
      keepaspectratio]{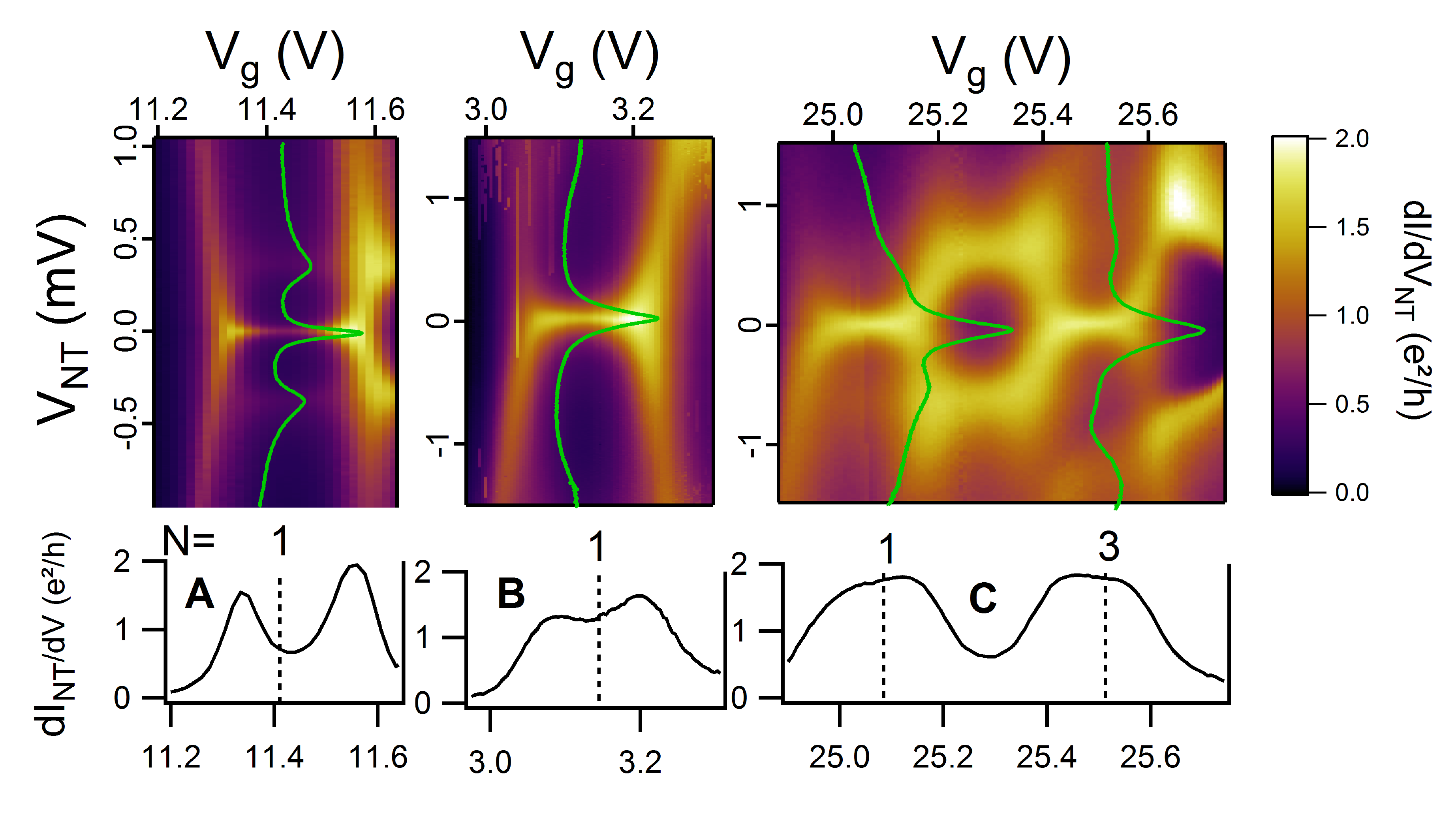}
    \end{center}
     \caption{Differential conductance $dI/dV_{NT}$ versus gate voltage $V_g$ and  bias voltage $V_{NT}$ for samples A, B and C. Green lines: $dI/dV_{NT}(V_{NT})$ at the center of each Kondo diamond. Below are represented horizontal cuts at zero bias voltage. Figure adapted from \cite{Delagrange2018}.}
     \label{NoiseHF_conductance}
\end{figure}

The differential conductance $dI/dV_{NT}$ of the CNT is measured, at very low frequency (below 100Hz) with respect to bias and gate voltage (Fig.\ref{NoiseHF_conductance}) for three samples, named A, B and C. They exhibit Coulomb diamonds with Kondo-induced zero-bias conductance peaks. Zone C includes two Kondo regions in two successive diamonds with odd number of electrons. The temperature dependence of the zero bias conductance peak allows to extract the Kondo temperature, and its height the asymmetry $a$ (see \cite{Delagrange2018} for A and C, and ref. \cite{Basset2012a} for B). This asymmetry is defined as $a=\Gamma_1/\Gamma_2$, with $\Gamma_{1,2}$ the coupling with the contacts 1, 2 and contact 1 being the most coupled one ($a\geqq 1$).  We get for zone A $T_K= 350mK$ and $a=11$, for zone B 1.4K and $a=5$, for zone C 1.5K and $a=1.5$ for both diamonds. For zones A and C, two satellite peaks are also visible, at $e V_{NT}=\pm0.35\mathrm{~meV}$ and $\pm0.6\mathrm{~meV}$, indicating a breaking of the CNT orbital degeneracy \cite{Laird2015,Makarovski2007a}.

\subsection{High frequency noise measurement in the SU(2) Kondo regime}
Simultaneously with conductance measurements, we probed emission noise in the center of the Kondo ridge : we measured, with a lock-in technique, the derivative of the PAT current in the detector versus the CNT bias voltage, modulated at low frequency (below 100Hz). This is done for two bias voltages of the detector, to extract the modulated PAT current corresponding to emission noise at the first two  resonance frequencies of the coupling circuit. This quantity is proportional to the derivative of the noise versus $V_{NT}$, $dS_I/dV_{NT}$ \cite{Delagrange2018}.

\begin{figure}[h]
     \begin{center}
    \includegraphics[%
      width=0.9\linewidth,
      keepaspectratio]{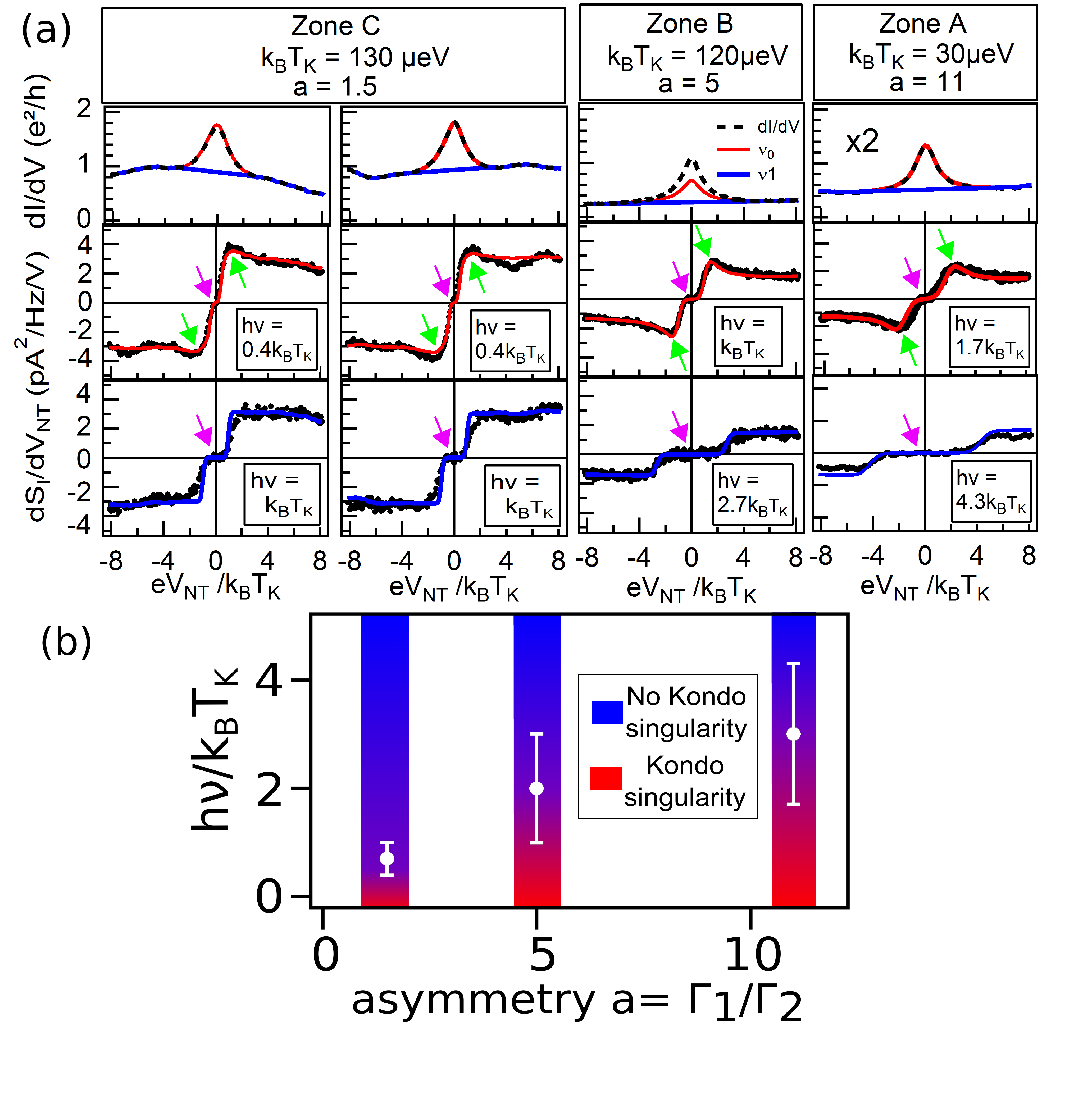}
    \end{center}
     \caption{(a) Comparison of the experimental data with the predicted $dS_I/dV_{NT}$ (see text) versus $eV_{NT}/k_BT_K$  for ridges A, B and C. Top panels : $dI/dV_{NT}$ (black dashed line) and effective conductance for the noise best fit at $\nu_0$ (red line, 12GHz for A and C, 29.5GHz for B) and at $\nu_1$ (blue line, 31GHz for A and C, 78GHz for B). For zone A the curves are scaled by a factor 2. Middle panels : $dS_I/dV_{NT}$ at $\nu_0$ (black dots) and calculated noise using $T(\epsilon)$ corresponding to the effective conductance shown in the top panel (red line). Bottom panels : same quantities at $\nu_1$ (black dots and blue line). (b) and (d) Schematic of the DOS of the quantum dot forming a Kondo singlet in the asymmetric and symmetric case. (c) Region of parameters where a Kondo related noise singularity is observed (red) or not measured (blue). This defines a range for the frequency cut-off (white dot). Figure adapted from \cite{Delagrange2018}.}
     \label{NoiseHF_noise}
\end{figure}
      
The noise measurements for sample A, B and C are presented in fig. \ref{NoiseHF_noise} where $V_{NT}$ has been rescaled by $T_K$, to emphasize the various $h\nu/k_BT_K$ ratios. Those curves exhibit two main features. The first one is a plateau centered around $V_{NT}=0$ with $dS_I/dV_{NT}=0$ (pink arrows in fig.\ref{NoiseHF_noise}). This is because we measure only emission noise, which is non-zero only for $V_{NT}>h\nu/e$. The second feature is a peak in the noise derivative at $V_{NT}=h \nu/e$, at the lowest frequency (green arrows in fig. \ref{NoiseHF_noise}). This peak is suppressed at higher frequency. We want now to compare the data with theoretical predictions.

There exist renormalisation group theories predicting the high frequency noise expected in the Kondo regime \cite{Moca2011,Muller2013} but they do not take into account the conductance background, present in our data, which is not related to the Kondo effect. That is why we rather compare the measurements to the noise calculated from the energy dependence of the transmission coefficient \cite{Zamoum2016,Hammer2011,Rothstein2009}. This quantity can be seen as the noise expected in a quantum dot that exhibits the same differential conductance $dI/dV$ as the one measured in our experiment, from which is extracted the energy dependent transmission $T(\epsilon)$. Recent work \cite{Crepieux2017} takes into account the Coulomb interaction in the dot and the asymmetry of the contacts and calculates the high frequency emission noise. It reproduces well the peak in the noise derivative at the lowest frequency.
 
The delicate point of our analysis is the extraction of the energy dependent transmission from the conductance measurement to account for the Kondo resonance. We assume that the energy dependent transmission in the Kondo regime is related to $dI/dV_{NT}$ \cite{Delagrange2018} by :
\begin{equation}
	T(\epsilon)=\frac{h}{2e^2} \frac{dI}{dV_{NT}}(\epsilon/e)
\label{transmission}
\end{equation}

The experimental data are compared with the calculated noise, using the energy dependent transmission coefficient extracted from the differential conductance (formula \ref{transmission}) and the expressions of the current noise given in ref. \cite{Zamoum2016, Delagrange2018}. This procedure is not able to explain the decrease at high frequency of the Kondo peak in the derivative of the noise. To quantify this disagreement, we have found, for each measurement, the value of the transmission coefficient that fits best our data. To do that, the amplitude of the Kondo peak close to zero bias is chosen as the fitting parameter while the baseline coming from the conductance background is kept unchanged. The effective conductance (given by formula \ref{transmission}) corresponding to this new transmission coefficient is shown in figure \ref{NoiseHF_noise}. For the lowest frequencies, we find that the effective conductance is similar or only slightly reduced compared to the real one. However, at higher frequency, it shows no signature of the Kondo resonance. 

These results show that, in our three samples, there exists a frequency cut-off above which the signature of the Kondo resonance in the emission noise vanishes. This cut-off is found to be around $k_B T_K/h$ in the symmetric case (sample C), when the voltage bias applied on the sample effectively drives the Kondo resonance out-of-equilibrium. The vanishing of the Kondo feature in the noise could then be attributed to voltage induced spin-relaxation, with a relaxation rate roughly proportional to bias voltage and $\Gamma_1 \Gamma_2/(\Gamma_1+\Gamma_2)^2$ \cite{Kaminski99,Kaminski00,Paaske04,Basset2012a,Muller2013}. However, according to ref. \cite{Leturcq2005} who measured the out-of-equilibrium density of states of a quite symmetric Kondo effect ($a=1.5$), a bias voltage of $V=4.5 k_B T_K/e$ is found to split the Kondo resonance into two peaks centered at $\pm eV/2$, without destroying it. The cut-off we measure cannot thus be accounted for by the effect of a bias voltage on the DOS of the system, and may be related to the high frequency nature of the measurement. This hypothesis is supported by the fact that we measure as well a frequency cut-off in the asymmetric cases (a=5 and a=11), although at a higher value (around $2-4 k_B T_K/h$). Indeed, for a strong contact asymmetry, one expects that the Kondo state formed with the best coupled contact stays very close to equilibrium at the chemical potential of this contact and is only slightly perturbed by the less coupled contact on which noise is measured. Consequently, we attribute the reduction of the Kondo feature in the emission noise mainly to dynamical effects, rather than out-of-equilibrium decoherence.

\subsection{High frequency noise measurement in the SU(4) Kondo regime}
\begin{figure}[htb]
      \begin{center}
      \includegraphics[height=9cm]{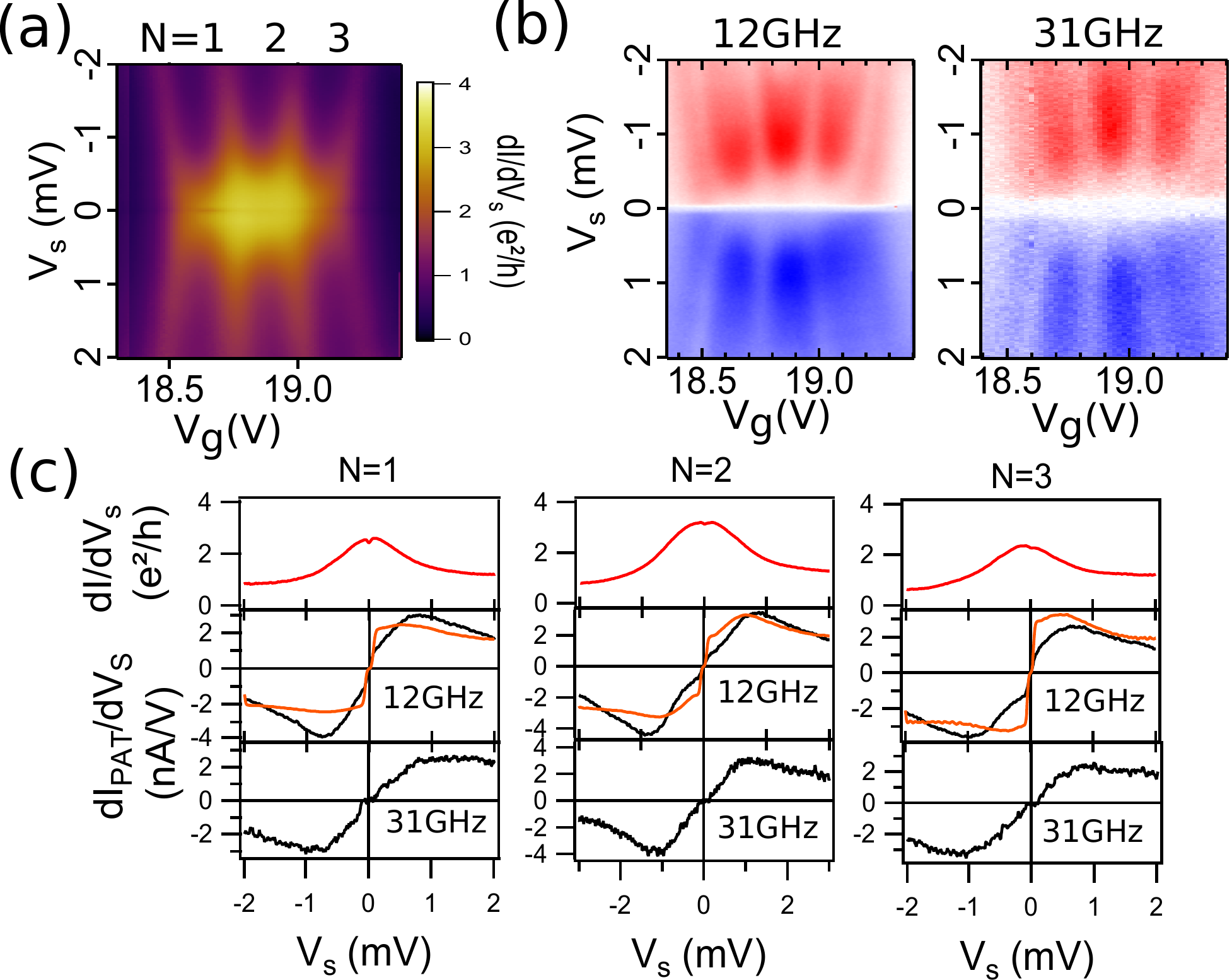}
      \end{center}
      \caption{Noise measurement in the SU(4) Kondo regime. (a) Differential conductance as a function of bias voltage $V_s$ and gate voltage $V_g$ in a region where the coupling of the QD to the electrodes is large enough to allow SU(4) Kondo effect(b) Derivative of the photo-assisted noise in the detector, proportional to the noise derivative, as a function of $V_s$ and $V_g$, in the Kondo SU(4) region. (c) Vertical cuts at the centers of the N=1, 2 and 3 diamonds of each of the three color plots.}
      \label{NoiseHF_su4}
\end{figure}

We also find a zone in gate voltage where the coupling between the CNT dot corresponding to sample C and the reservoirs is large enough to observe SU(4) Kondo effect. The conductance map is represented in fig. \ref{NoiseHF_su4}. 

As for SU(2) Kondo effect, we measure the noise derivative as a function of the bias $V_s$ and gate voltage $V_g$, all over the three diamonds of the SU(4) Kondo region ($m=$1, 2 and 3). The frequencies at which the noise is measured are 12 GHz and 31 GHz, but the Kondo temperature here is of the order of $500\mathrm{~\mu eV}$ ($T_K\approx6\mathrm{~K}$), corresponding to 125 GHz. The measurement is thus in a low frequency regime such that, except the width of the emission noise plateau at low voltage, the noise is the same at both frequencies.

Moreover, in this regime it is difficult to compare the experimental data with the scattering matrix theory. Indeed, since two channels are involved, the extraction from the conductance of the energy dependent transmission is more complex. We have tried to do it assuming that the two channels were equally transmitted (see orange lines in fig. \ref{NoiseHF_su4}), but the agreement with the experiment is very poor suggesting that this hypothesis is wrong.

\subsection{Conclusions for the high frequency noise measurement}

To conclude, we have measured the high frequency emission noise of a CNT QD in the Kondo regime with different symmetries (SU(2) and SU(4)) and for various coupling asymmetries of the reservoirs with the dot, in the SU(2) regime. In this latter case, at the lowest measured frequencies the derivative of the noise exhibits a Kondo peak, well reproduced by theories which compute the finite frequency noise from the energy dependent transmission. This peak is strongly suppressed at higher frequency, pointing towards the existence of a high frequency cut-off of the electronic emission noise at a Kondo resonance. In the symmetric case, this cut-off can be partially accounted for by decoherence effects. However in the asymmetric case, the Kondo state is not driven out-of-equilibrium by the bias voltage. This leads us to postulate that in QDs in the Kondo regime a new timescale, related to the Kondo energy $k_B T_K$, emerges besides the natural timescale associated to the transport of electrons through the dot, given by $\Gamma$ the coupling to the reservoirs. This statement is however not supported by existing theories, which predict a very slow frequency dependence of ac properties in the Kondo regime \cite{Moca2011,Muller2013}. This motivates further investigations to understand better the role of the Kondo dynamics in the high frequency current fluctuations.

\section{Conclusions}
In this paper, we have demonstrated that current noise measurement is a powerful tool to probe quantum dots in the Kondo regime, and carried out such measurement in two complementary limits.
At low frequency, we obtained a quantitative understanding of the role of electron-electron interaction. We have demonstrated that interactions emerge beyond equilibrium leading to increased quantum fluctuations and the appearance of a peculiar two-particle backscattering.

However, noise measurement in the quantum regime provides some puzzling results, which remain to be understood. It points towards the existence of a frequency cut-off on the current fluctuation in the Kondo regime. This cannot be only explained by decoherence effects and raised the question of the signature of the internal dynamics of the Kondo effect on current fluctuations.

\begin{acknowledgements}
The authors acknowledge fruitful discussions with A. Cr\'epieux, M. Lavagna, P. Simon, S. Gu\'eron, A. Chepelianskii and B. Reulet. This work was supported by the French programs ANR MASH (ANR-12-BS04-0016), DYMESYS (ANR 2011-IS04-001-01), DIRACFORMAG (ANR-14-CE32-0003), JETS (ANR-16-CE30-0029-01), the Japan Society for the Promotion of Science KAKENHI Grant No.~JP26220711, No.~JP18K03495,  No.~JP15K17680, No.~JP19H05826, No.~JP18J10205 and No.~JP19H00656, the JST CREST Grant No. JPMJCR1876 and the Yazaki Memorial Foundation for Science and Technology, Research Institute of Electrical Communication, Tohoku University.\\
\end{acknowledgements}

\newpage

\section*{Appendix}

\subsection*{\textbf{Wilson ratio: how to characterize interaction and fluctuations in a Fermi-liquid}}
\label{wilson}
A key parameter for Fermi-liquid theory is the Wilson ratio. It is defined as the ratio between the spin susceptibility to the linear part of the temperature-dependent specific heat:
\begin{equation}
R=\frac43\left(\frac{\pi k_B}{g\mu_B}\right)^2\frac{\chi}{c_V/T}
\end{equation} 
This definition removes the contribution of the effective mass which enters both in $\chi$ and $c_V$. Thus, $R$ quantifies spin fluctuations that enhance the susceptibility.
For a Fermi-liquid this ratio is a constant directly related to the usual interaction parameter $F$ ($<0\ $ for repulsive interaction) of the Landau theory\cite{Zou1986}:
\begin{equation*}
R=\frac{1}{1+F}
\end{equation*}

For a free electron gas we obtain $R=1$ whereas it increases to the universal value $R=2$ for any system in the spin $1/2$ (SU(2)) Kondo regime. This also reminds that spin fluctuations of the Kondo ground state are at the heart of the residual interaction of the quasiparticles forming the surrounding Fermi-liquid as explained by Nozieres.

In a quantum dot coupled to electrodes, $R$ can be computed for any value of $\frac{U}{\Gamma}$. The result is shown in Fig. \ref{oguriTk}, where we can see that $R\rightarrow 2$ when the dot enters the Kondo regime $\frac{U}{\Gamma}\rightarrow \infty$.

Actually, \textbf{at half filling}, this ratio and thus spin quantum fluctuations depend on the symmetry $SU(N)$ of the Kondo state with the relation:
\begin{equation}
R=1+\frac{1}{N-1}
\end{equation}

This means that the amplitude of spin fluctuations between two components of the total spin\footnote{In the well defined SU(N) symmetry, all components of the total spin are equivalent. But in the case of broken symmetry different $R$ can be defined to characterize fluctuations between each components \cite{Teratani2016}}  (and the associated residual interaction) decreases when the number of degree of freedom of the local impurity increases. 

 For SU(N), the number of spin states on the dot increases with $N$ and the relative change induced by a fluctuation is less and less important. In the limit $N\rightarrow\infty$, we recover the mean field result without fluctuations. The amplitude of fluctuations reflects the "discreteness" of the local moment. It is also a measurement of the deviation from the mean field approximation.


\subsection*{\textbf{Fermi coefficients and Kondo temperature}}
Around equilibrium, in the linear regime, transport properties can be described simply in the framework of the Landauer Buttiker theory of non-interacting particles. This is no longer the case in the non equilibrium regime and non linear terms appear in the current and noise due to interaction between the Landau quasi-particles.
This statement is general for any interacting system. Remarkably, Kondo effect is expected to exhibit universal behaviour even out of equilibrium. This makes it a very precise system to check general predictions for Fermi liquids beyond the well known equilibrium situation.

\begin{figure}[h] 
  \centering 
  \includegraphics[width=0.6\columnwidth]{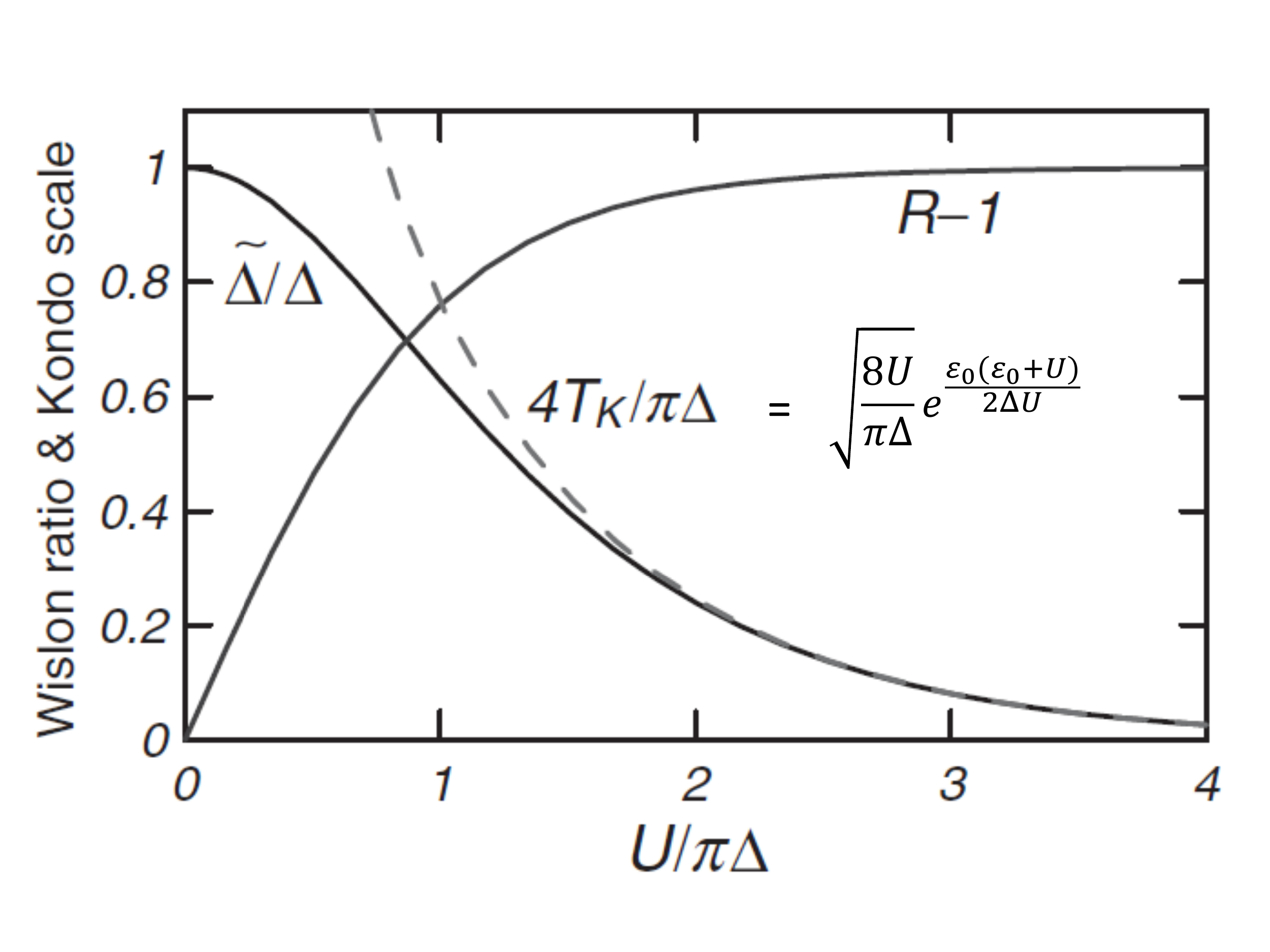}

  \caption{Evolution of R and $\tilde{\Delta}$ as a function of $U/\Gamma$. The dotted line represents Eq.$3$ in the main text. $T_K$ can be calculated from this equation when it is in agreement with $\tilde{\Delta}$, which happens in the strong interaction regime called Kondo regime. Taken from \cite{Oguri2005}. In this figure $\pi\Delta=\Gamma$ in our notations.
  }
  \label{oguriTk} 
  \end{figure}

Taking interaction into account, the current has been derived until the third order ($T^2V$,$B^2V$ and $V^3$ for SU(2))\cite{Oguri2001,Pustilnik2004}. Results are given by a single energy scale $T_K$ only in the "Kondo" limit $U/\Gamma\gg 1$. In the intermediate regime, NRG computation is necessary, yielding for the conductance:
\begin{equation}
G(T,B,V_{sd})=G(0)\left[ 1-c_T\left(\frac{k_BT}{\tilde{\Delta}}\right)^2-c_B\left( \frac{g\mu_BB}{\tilde{\Delta}}\right) ^2-c_V\left( \frac{eV}{\tilde{\Delta}}\right) ^2\right] 
\label{Gscaling}
\end{equation}
 where the Fermi-liquid coefficients $c_T, c_B$ and $c_V$ depends only on the Wilson ratio $R$ through:
 \begin{equation}
 c_T=\pi^2\frac{1+2(R-1)^2}{3},\quad c_B=\frac{R^2}{4},\quad  c_V=\frac{1+5(R-1)^2}{4}.
 \label{Fermicoef}
 \end{equation}
   $\tilde{\Delta}$ is  the renormalized width of the Kondo resonance. It depends on the parameters $U$ and $\Gamma$ and it is only in the strongly  interacting limit that $\tilde{\Delta} \rightarrow \frac{4T_K}{\pi}$ where $T_K$ is given by the exponential law of Eqn. $3$ in the main text and\cite{Oguri2005} $R=2$. This can be seen on Fig.\ref{oguriTk}, which represents $\tilde{\Delta}$ and $T_K$ from Eq.$3$ of main text as a function of $\frac{U}{\Gamma}$. A reasonable agreement is already achieved for $\frac{U}{\Gamma}\geq2$.
 
 In our first experiments, we have used these coefficient to determine self-consistently $R$ and $T_K$ in a CNT dot in the SU(2) Kondo state (see also supplementary material in \cite{Ferrier2015a}).
 
 These coefficients have been also extended in the general $SU(N)$ case\cite{Horig2014,Hanl2014}. However we couldn't check it experimentally with a good accuracy. The Kondo temperature in the SU(4) case was so high ($T_K>10\ $K) that variations of $G$ were too weak to extract these coefficients.
 
 \subsection*{\textbf{Extraction of Fermi coefficients and Kondo temperature}} 
 This analysis has been performed on ridges $C$ and $D$ of Fig.~5 in the main text, which present SU(2) symmetry with $G=0.85G_Q$ and $G=G_Q$. This high conductances denote a very good left/right symmetry and allow a precise quantitative analysis of transport. We have developed a method to evaluate $T_K$ and Wilson ratio $R$ without parameter from the Fermi coefficients(see supplementary in \cite{Ferrier2015a} ).
  We just rewrite Eq.\ref{Gscaling} as:
  \begin{equation*}
  G(T,B,V_{sd})=G(0)\left[ 1-\alpha_TT^2-\alpha_B(g\mu_BB) ^2-\alpha_V(eV) ^2\right] 
  \end{equation*}
  
  The ratio of two parameters ($\alpha_B$ and $\alpha_V$ are more precise than $\alpha_T$ in our experiments) yields the Wilson ratio through $\frac{\alpha_V}{\alpha_B}=\frac{c_V}{c_B}=\frac{1+5\left(R-1\right)^2}{R^2}$. $c_V$ and $c_B$ are defined in Eq. \ref{Fermicoef}. 
  
   Since we found $R\sim 1.7$ and $R\sim 1.95$ for the ridges $C$ and $D$, the strong coupling limit is achieved ($U/\Gamma\gg 1$). Indeed, from Fig. \ref{oguriTk} we can see that for this values of $R$, the curves of $T_K$ and $\tilde{\Delta}$ are superimposed, which ensures that the substitution $\tilde{\Delta} \rightarrow \frac{4T_K}{\pi}$ is allowed. Thus, combining Eq.\ref{Gscaling} and \ref{Fermicoef}, we obtain the expression of $\alpha_B$ or $\alpha_V$ a function of $T_K$. For example $\alpha_B=\left(\frac{\pi R}{8k_BT_K}\right)^2$ and we could safely extract $T_K$ from the value of $\alpha_B$.
  
  Using this result for $T_K$ we have confirmed that the temperature dependence of the conductance follows the phenomenological law: 
  \begin{equation}
   G(T)=G(0)\left(1+(2^{1/s}-1)\left(\frac{T}{T_K}\right)^2\right)^{-s}
   \label{GdeT}
   \end{equation}
  By fitting the Temperature dependence of G with $s$ as a single parameter we have determined $s=0.22\pm 0.005$ in perfect agreement with NRG and previous experiments in the SU(2) regime \cite{Kretinin2011}. Again in he SU(4) case, due to the high $T_K$ it was not possible to measure accurately the temperature dependence of the conductance.
  
  \subsection*{\textbf{NRG model}}
  
The NRG computations were realized by Akira Oguri and Yoshimichi Teratani from Osaka city university.
  
  In our calculations the CNT dot is modelled by an Anderson impurity coupled to two leads labelled by $\alpha=L,R$ with the Hamiltonian $H=H_d+H_T+H_c$ :
  \begin{eqnarray*}
  H_d & = & \sum_{m=1}^{4}\epsilon_mn_m+U\sum_{m<m'}n_mn_{m'}\\
  H_T & = & \sum_{\alpha=L,R}\sum_{m=1}^{4}\int_{-D}^{D}d\epsilon\sqrt{\rho_c}v_\alpha c^{\dagger}_{\alpha,\epsilon_m}d_m+h.c \\
  H_c & = &\sum_{\alpha=L,R}\sum_{m=1}^{4}\int_{-D}^{D}d\epsilon\epsilon c^{\dagger}_{\alpha,\epsilon_m}c_{\alpha,\epsilon_m}
  \end{eqnarray*}
  $H_d$ is the dot Hamiltonian, $H_c$ the conducting electrons in the leads, and $H_T$ represents the coupling between the dot and leads.
  The four dot states are labelled by $m=1, 2, 3, \mbox{and~} 4$ as defined in Fig.~1a in the main text. We note $n_m=d_m^{\dagger} d_m$ and $\rho_c=1/(2D)$ with $D$ the half bandwidth of the conduction band. The level width is given by $\Gamma\equiv 2\Delta=4\pi\rho_c v^2$, where we assume symmetric couplings $v_L=v_R =v$. NRG calculations are carried out for $U=3.15\Gamma$, taking discretization parameter $\Lambda=6.0.$ We have kept typically $3,000$ states in each NRG step using the $U(1)\times U(1)\times U(1) \times U(1)$ symmetry corresponding to charge conservation for each channel $m$.\\
  
  The dot states consist of  the spin ($\uparrow$, \!$\downarrow$) 
  and valley  ($\mathrm{K}$, \!\! $\mathrm{K'}$) degrees of freedom. In the presence of magnetic field, the one-particle energies $\epsilon_m$ are the eigenvalues 
  of Hamiltonian $H_0^\mathrm{one}$  
   which takes the following form with a basis set of 
   $|\mathrm{K} \uparrow \rangle$, 
   $|\mathrm{K} \downarrow \rangle$, 
   $|\mathrm{K}'\uparrow \rangle$, and 
   $|\mathrm{K}'\downarrow \rangle $:  
   
   \begin{align*}
   \!\!\!\!\!\!\!\!\!\!
   H_0^\mathrm{one}
    = 
   \begin{psmallmatrix} 
   \epsilon_d +\frac{\Delta_\mathrm{SO}}{2} 
   + ( g_\mathrm{orb} + \frac{g_\mathrm{s}}{2} ) \mu_B B_{\parallel}^{}
   & \frac{g_\mathrm{s}}{2}\, \mu_B B_{\perp}^{} & \frac{\Delta_\mathrm{KK'}}{2}& 0  
   \cr 
    \frac{g_\mathrm{s}}{2} \,\mu_B B_{\perp}^{} 
   & \epsilon_d -\frac{\Delta_\mathrm{SO}}{2}  
   +( g_\mathrm{orb} - \frac{g_\mathrm{s}}{2})  \mu_B B_{\parallel}^{}
   & 0 & \frac{\Delta_\mathrm{KK'}}{2} 
   \cr 
     \frac{\Delta_\mathrm{KK'}}{2} & 0 & 
   \epsilon_d - \frac{\Delta_\mathrm{SO}}{2} 
   -( g_\mathrm{orb} - \frac{g_\mathrm{s}}{2} ) \mu_B B_{\parallel}^{}
   & \frac{g_\mathrm{s}}{2} \mu_B B_{\perp}^{}  
   \cr 
    0 & \frac{\Delta_\mathrm{KK'}}{2}  & \frac{g_\mathrm{s}}{2} \mu_B B_{\perp}^{} &
   \epsilon_d + \frac{\Delta_\mathrm{SO}}{2}  
   -( g_\mathrm{orb} + \frac{g_\mathrm{s}}{2}) \mu_B B_{\parallel}^{}
   \cr 
   \end{psmallmatrix}         
   \end{align*}
   
  Here,  $B_{\parallel}^{} \equiv  B \cos{\theta}$, $B_{\perp}^{} \equiv  B \sin{\theta}$, and $\theta$ is the angle of the magnetic field relative to the nanotube axis. $\Delta_{SO}$ is the spin-orbit coupling, and $\Delta_{KK'}$ the intervalley scattering. 
  
  \subsubsection*{Effect of the Magnetic Field and Determination of the angle}
  
  In the experiment, we have first extracted the angle of the magnetic field from the field evolution of excited states of the CNT dot. Then, we have confirmed the precise value for the angle by comparison with NRG computation at different values for $\theta$. In the following, we describe this procedure in more detail.
  
  \subsubsection*{Magnetospectroscopy of the excited states}
\begin{figure}[tb]
  \center
  \includegraphics[width=0.85\linewidth]{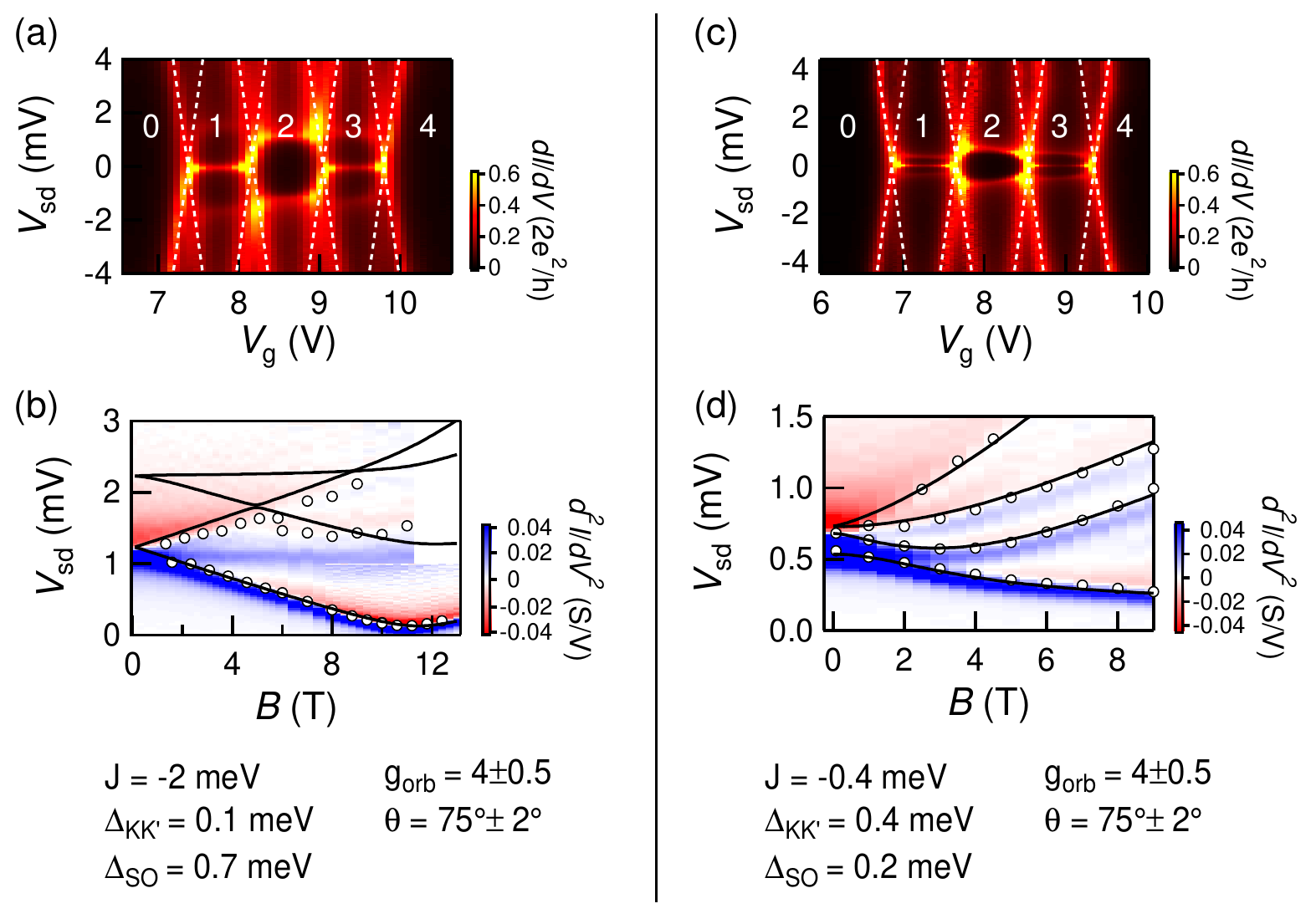}
  \caption{ \textbf{(a)} Stability diagram at the gate voltages different from those shown in the main text. \textbf{(b)}  Fit of inelastic co-tunnelling peaks on the shell. The colour plots represent the second derivative of the current $\frac{d^2I}{dV^2}$. Hence, when increasing $V_{sd}$, a peak in the conductance corresponds to zero preceded by negative values and followed by positive values. These are white lines surrounded by blue area below and red area above. Symbols mark the position of experimental peaks in the conductance. Dotted line is the result fit. The fitted parameters are shown in the bottom. \textbf{(c)} and \textbf{(d)} The counterpart of (a) and (b) for other shell. As the gate voltage has shifted between these two experiments, the diagrams (a) and (c) correspond to different shells. }
  \label{Fitmagneto}
  \end{figure}

  We measured the stability diagrams at gate voltages different from those of the main text and selected two distinct shells displayed in  Figs. \ref{Fitmagneto}a and c. In these shells, at $N=2$, due to the exchange energy, the spin-orbit coupling,  and the intervalley scattering the eigenstates at $B=0$ are not degenerate. The large splitting prevents the formation of Kondo resonance. Consequently, the inelastic co-tunnelling peaks, which correspond to transitions between the ground state and the excited states appear in the stability diagram at finite $V_{sd}$~\cite{Cleuziou2013}.
  
  The position of the eigenstates depends on $B$ and was calculated from the following Hamiltonian $ H_{0}^\mathrm{two}+ H_{1}^\mathrm{two}$ describing two electrons states in the CNT with an in-plane magnetic field which forms an angle $\theta$ with the CNT axes.
  
  {\small
  \[
    H_{0}^\mathrm{two} = \left(
  \begin{array}{cccccc}
        2\varepsilon_{\rm \scalebox{0.5}{d}}+2g_{\rm \scalebox{0.5}{orb}}\mu_{\rm \scalebox{0.5}{B}}B\rm{cos}\theta & 0 & \dfrac{1}{\sqrt{2}}\Delta_{\rm \scalebox{0.5}{KK'}} & 0 & 0 & 0 \\
        0 & 2\varepsilon_{\rm \scalebox{0.5}{d}}-2g_{\rm \scalebox{0.5}{orb}}\mu_{\rm \scalebox{0.5}{B}}B\rm{cos}\theta & \dfrac{1}{\sqrt{2}}\Delta_{\rm \scalebox{0.5}{KK'}} & 0 & 0 & 0 \\
        \dfrac{1}{\sqrt{2}}\Delta_{\rm \scalebox{0.5}{KK'}} & \dfrac{1}{\sqrt{2}}\Delta_{\rm \scalebox{0.5}{KK'}} & 2\varepsilon_{\rm \scalebox{0.5}{d}} & \Delta_{\rm \scalebox{0.5}{so}} & 0 & 0 \\
        0 & 0 & \Delta_{\rm \scalebox{0.5}{so}} & 2\varepsilon_{\rm \scalebox{0.5}{d}} & \dfrac{1}{\sqrt{2}}g_{\rm \scalebox{0.5}{s}}\mu_{\rm \scalebox{0.5}{B}}B\rm{sin}\theta & \dfrac{1}{\sqrt{2}}g_{\rm \scalebox{0.5}{s}}\mu_{\rm \scalebox{0.5}{B}}B\rm{sin}\theta \\
        0 & 0 & 0 & \dfrac{1}{\sqrt{2}}g_{\rm \scalebox{0.5}{s}}\mu_{\rm \scalebox{0.5}{B}}B\rm{sin}\theta & 2\varepsilon_{\rm \scalebox{0.5}{d}}+g_{\rm \scalebox{0.5}{s}}\mu_{\rm \scalebox{0.5}{B}}B\rm{cos}\theta & 0 \\
        0 & 0 & 0 & \dfrac{1}{\sqrt{2}}g_{\rm \scalebox{0.5}{s}}\mu_{\rm \scalebox{0.5}{B}}B\rm{sin}\theta & 0 & 2\varepsilon_{\rm \scalebox{0.5}{d}}-g_{\rm \scalebox{0.5}{s}}\mu_{\rm \scalebox{0.5}{B}}B\rm{cos}\theta
  \end{array}
    \right)
  \]
  }
  \\
  {\small
  \[
    H_{1}^\mathrm{two} = \left(
      \begin{array}{cccccc}
        U& 0  & 0  & 0  & 0  & 0 \\
        0  & U  & 0  & 0  & 0  & 0 \\
        0 &  0 &  \dfrac{1}{2}J_{\rm \scalebox{0.5}{KK'}}+U &  0  & 0  & 0 \\
        0 &  0 &  0 & -\dfrac{1}{2}J_{\rm \scalebox{0.5}{KK'}}+U & 0 & 0 \\
        0 & 0 & 0 & 0 & -\dfrac{1}{2}J_{\rm \scalebox{0.5}{KK'}}+U & 0 \\
        0 & 0 & 0 & 0 & 0 & -\dfrac{1}{2}J_{\rm \scalebox{0.5}{KK'}}+U
  \end{array}
    \right)
  \]
  }
  Here, we have used the two-particles basis set defined as follows:
  \begin{align*}
  &|\mathrm{A}\rangle\,=\,d_{K\uparrow}^{\dagger}d_{K\downarrow}^{\dagger}|0\rangle, \qquad
  |\mathrm{B}\rangle\,=\,d_{K^{\prime}\uparrow}^{\dagger}d_{K^{\prime}\downarrow}^{\dagger}|0\rangle ,
  \\
  & |\mathrm{C}\rangle\,=\,\frac{1}{\sqrt{2}}(d_{K\uparrow}^{\dagger}d_{K^{\prime}\downarrow}^{\dagger}\,-\,d_{K\downarrow}^{\dagger}d_{K^{\prime}\uparrow}^{\dagger})|0\rangle ,
  \qquad 
  |\mathrm{D}\rangle\,=\,\frac{1}{\sqrt{2}}(d_{K\uparrow}^{\dagger}d_{K^{\prime}\downarrow}^{\dagger}\,+\,d_{K\downarrow}^{\dagger}d_{K^{\prime}\uparrow}^{\dagger})|0\rangle ,\\
  &|\mathrm{E}\rangle\,=\,d_{K\uparrow}^{\dagger}d_{K^{\prime}\uparrow}^{\dagger}|0\rangle , 
  \qquad 
  |\mathrm{F}\rangle\,=\,d_{K\downarrow}^{\dagger}d_{K^{\prime}\downarrow}^{\dagger}|0\rangle. \\
  \end{align*}
  
  Since the conductance peaks occur at voltages $eV_{sd}=E_i-E_0$, where $0$ indicates the ground state and $i$ labels the excited states, we have numerically diagonalized this Hamiltonian and fit the position of the peaks in the conductance with five parameters ($g_{orb}$, $\theta$, $J_{KK'}$, $\Delta_{SO}$, and $\Delta_{KK'}$). ($\epsilon_d$ and $U$ only shift constantly the energies and do not play any role here). For the fit, we impose that $g_{orb}$ and $\theta$ are the same for both shells. The three other parameters are independent since it has been shown that they can depend on the shell~\cite{Cleuziou2013, Sapmaz2005}. The result is shown in Fig\ref{Fitmagneto}. The number of different excitation lines, especially the anti-crossing at high field, gives a reasonable accuracy to the fitting parameters. We have obtained $g_{orb}=4\pm 0.5$ and $\theta = 75^{\circ} \pm 2$, which result in $g_{orb}\cos\theta\approx 1$. This result is reasonable since we expect\cite{Laird2014} $g_{orb}\approx 3.5d$ where $d$ is the diameter of the CNT in nm  which is $d\approx 1\ $nm in our case. Also we have mounted the sample in the dilution fridge such that the field is in plane, nearly perpendicular to the contacts. Other parameters are displayed in the the bottom of Figs.~\ref{Fitmagneto}.
  
  \subsubsection*{Confirmation by NRG computation.}
  
\begin{figure}[tb]
  \centering
  \includegraphics[width=\linewidth]{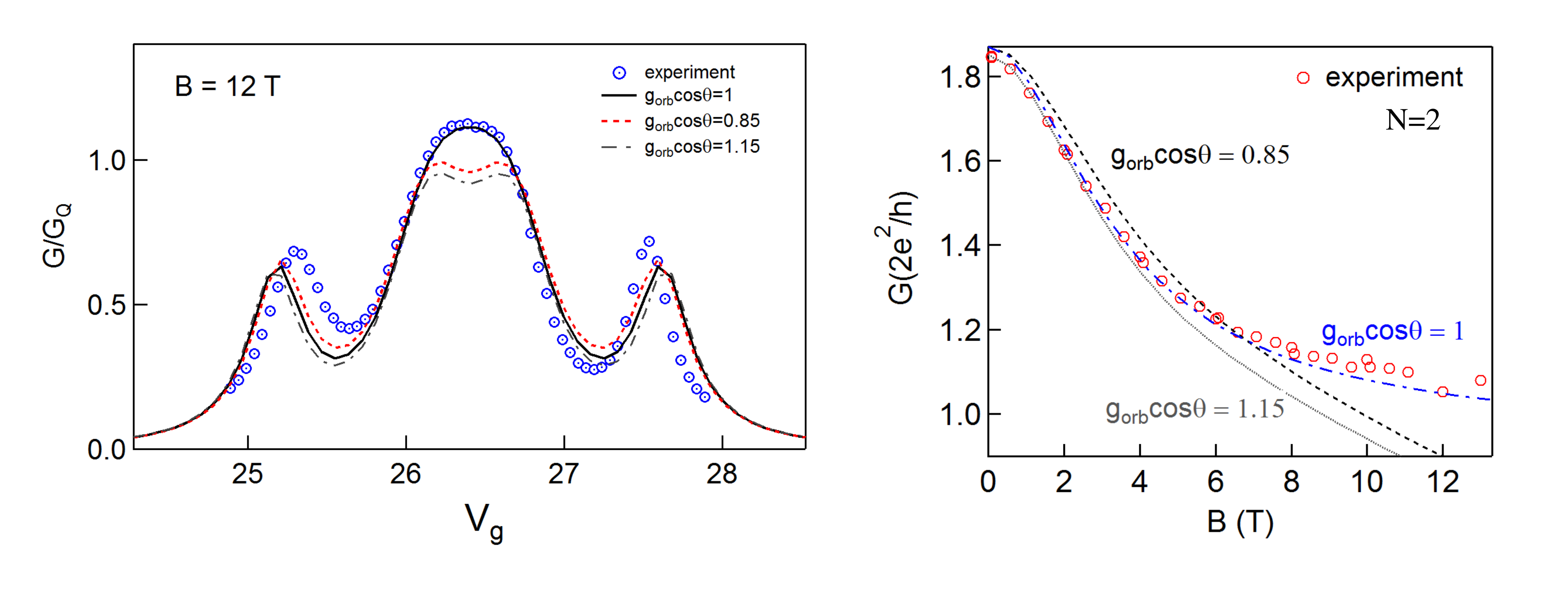}
  \caption{Evaluation of the parameter $g_{orb}\cos{\theta}$ from NRG computation. \textbf{Left)} Conductance as a function of $V_g$ in the $SU(2)$ limit of the crossover (12~T). Symbols are the experimental data. The lines are NRG results for  $g_{orb}\cos{\theta}=0.85, 1$, and 1.15. \textbf{Right)} Conductance as a function of magnetic field for the filling $N=2$ ($V_g\approx 26.4\ $V on the left panel). Lines represent NRG results for the same values of $g_{orb}\cos{\theta}$. The agreement between experiment and theory is clearly the best when $g_{orb}\cos{\theta}=1$ is assumed and allows us to assume $g_{orb}\cos{\theta}=1\pm 0.1$.}
  \label{Comp0p7}
  \end{figure}

  Once we evaluated $g_{orb}\cos\theta$, we obtained confirmation with NRG computation. Coming back to the shell of the main text, which presents the crossover at $N=2$, we have computed $G$ as a function of field for different values of $\theta$. As explained in the main text and shown in Fig.\ref{Comp0p7} there, NRG results reproduce experimental curves $G(V_g)$ at every magnetic field assuming $g_{orb}\cos\theta=1$. To check the accuracy, the left panel of Fig.~\ref{Comp0p7} presents the experimental curve $G(V_g)$ at $12\ $T compared with NRG results at the same field with different values for $g_{orb}\cos\theta$. It clearly shows that $g_{orb}\cos\theta=1$ yields the best agreement. The right panel of Fig.~\ref{Comp0p7} represents the magneto-conductance at $N=2$ compared with NRG results for different values of $g_{orb}\cos\theta$. Here also the best agreement is obtained for $g_{orb}\cos\theta=1$. This confirms that it is reasonable to assume $g\cos\theta = 1\pm0.1$.

  \subsubsection*{Robustness of the crossover}
  \label{robustness}
  Although the condition of $g_{orb}\cos{\theta}=1$ makes the explanation of the crossover easier to understand, it is not a stringent condition to observe it. Indeed, we checked carefully using our NRG model that this crossover occurs in an extended range of values for $g_{orb}\cos{\theta}$ around $1\pm 0.2$. We also incorporate $\Delta_{so}$ and $\Delta_{KK'}$ up to $0.2\ $meV. This is the maximum possible value since it determines $G$ at $B=0\ $T. For higher values of $\Delta_{so}$ and $\Delta_{KK'}$ the theoretical conductance becomes smaller than the measured one at $B=0$. Moreover the shape of $G(V_g)$ starts to strongly differ from the experimental one. In all  cases, the Wilson ratio which is the signature of the Kondo correlations, always crosses from a value close to the SU(4) limit ($R=4/3$) at low field to a value close to the SU(2) limit ($R=2$) at high field. The conductance at $N=2$ also decreases from $G\approx 2G_Q$ to $G\approx G_Q$ as in the experiment.
  Actually, the important point is that the splitting between the two lowest eigenstates of the two-particle Hamiltonian ($N=2$) is smaller than $k_BT_K$. Since the sample presents a large $T_K$ it makes this crossover possible for a reasonable range of values for $\theta$. The theoretical aspect regarding this topic is presented in more details in Ref.~\cite{Teratani2016}.

\end{document}